\documentclass[12pt]{article}
\usepackage[margin=0.5in]{geometry}
\usepackage{array,multirow}
\usepackage{verbatim}
\usepackage{graphicx}
\usepackage{bbm}
\usepackage{natbib}
\usepackage{algorithm,algorithmicx,algpseudocode}
\algdef{SE}[FORUNTIL]{for}{until}[1]{\algorithmicfor \ #1 \algorithmicdo}[1]{\algorithmicuntil\ #1}%
\usepackage{amsmath,amssymb,amsfonts,graphicx,nicefrac,mathtools,bm}
\usepackage{amsthm}
\usepackage{hyperref}
\usepackage{color}
\usepackage{xspace}
\newtheorem{theorem}{Theorem}

\newtheorem{lemma}{Lemma}

\newtheorem{condition}{Condition}
\begin{document}

\title{On High Dimensional Covariate Adjustment for Estimating Causal Effects in Randomized Trials with Survival Outcomes}

\author{Ran Dai \thanks{Department of Biostatistics, University of Nebraska Medical center},
        Cheng Zheng \thanks{Department of Biostatistics, University of Nebraska Medical center},
        Mei-Jie Zhang \thanks{Division of Biostatistics, Medical College of Wisconsin}}


\maketitle

\begin{abstract}
The purpose of this work is to improve the efficiency in estimating the average causal effect (ACE) on the survival scale where right-censoring exists and high-dimensional covariate information is available. We propose new estimators using regularized survival regression and survival random forests (SRF) to make the adjustment for the high dimensional covariates to improve efficiency. We study the behavior of the adjusted estimator under mild assumptions and show theoretical guarantees that the proposed estimators are more efficient than the unadjusted ones asymptotically when using SRF for adjustment. In addition, these adjusted estimators are $\sqrt{n}$- consistent and asymptotically normally distributed. The finite sample behavior of our methods are studied by simulation, and the results are in agreement with the theoretical results. We also illustrate our methods by analyzing the real data from transplant research to identify the relative effectiveness of identical sibling donors compared to unrelated donors with the adjustment of cytogenetic abnormalities.
\end{abstract}

\section{Introduction}
\label{intro}

Survival analysis is widely used in the assessment of interventions in clinical trials. In order to provide guidance on future interventions or treatment plans, the estimation of the causal effect is highly desirable. Double-blinded randomization trials are widely used to estimate the average causal effect (ACE) \citep{chalmers1981}. The crude unadjusted estimation of ACE is unbiased \citep{rubin1974}. However, as pointed out by \citet{fisher1925}, adjustment of (low dimensional) features that are correlated with the outcome using an ordinary least square method can reduce the variance of the estimated treatment effect for randomized studies without introducing any bias. The theoretical results are complete for low dimensional covariate cases \citep{rosenbaum1987,rosenbaum2002,freedman2008,cole2004,lin2014}. 
\paragraph{}
The adjustment using high dimensional covariates in treatment effect estimation in randomized survival trials was motivated for several reasons. First, the explosion of data made this type of adjustment possible. For example, in many clinical trials, high dimensional baseline information such as genomics, metabolomics or proteomics data are available that can potentially be used as covariates for the adjustment. Second, the relationships between the covariates and the response are complicated. For example, it is often the case that the covariates related to the response are correlated with possible interactions. Third, methods such as penalized survival regression model \citep{tibshirani1997} and survival random forest \citep{ishwaran2008,ishwaran2010} are developed to deal with high dimensional covariates in survival analysis, providing opportunities for high-dimensional covariate adjustment.
\paragraph{}
However, studies in using the high dimensional covariate adjustment to improve the efficiency of the estimation of ACE only emerged recently and is only limited for linear models. \citet{bloniarz2016} proposed adjustment with lasso, which, under a relatively strong sparsity assumption improves efficiency in the estimation of the ACE. For observational study, \citet{belloni2014} proposed a de-biased estimator, which allows inference for ACE after adjusting for high-dimensional potential confounders, at the price of widening the confidence interval. More recently, \citet{lei2020} proposed a new de-biased estimator for high-dimensional inference under much milder assumptions and allows the number of covariates to diverge under finite population framework. Under super-population framework, \citet{wager2016} showed that any intercept-free adjustment can produce unbiased estimation of the ACE, and the estimation precision only depends on the prediction risk of the fitted regression adjustment. 

\paragraph{}
There is a gap in characterizing the high dimensional covariate adjustment in survival analysis with censored outcome. In this paper we propose inverse probability censoring weighting (IPCW) based adjusted estimators for the ACE. Our main contributions are:
\begin{itemize}
\item We propose four high-dimensional covariate adjustment estimators using IPCW technique.
\item We show the consistency and asymptotic variance reduction results of the proposed estimators. The assumptions needed are weak, so that the adjustment can be done by many regularized regression methods and nonparametric machine learning methods including the penalized survival regression and nonparametric survival random forest. 
\item We show the finite sample performance of our proposed methods with simulations and a real data example.
\end{itemize}


\paragraph{}

The remaining sections are structured as follow. In Section 2 we first define notations and propose our estimators. In section 3, we provide our theoretical results on the asymptotic distributions of proposed estimators and study their relative efficiency. In Section 4, we present the simulation results to study the finite sample property of different estimators. In Section 4, we apply our method to a real data set for illustration. In Section 5, we give more discussions on the methods.

\section{Method}
\label{sec:1}

\subsection{Notations}
\label{sec1subsec1}
For individual $i$, we denote the randomization indicator as $Z_i \sim Z \in \{0,1\}$, where $\mathbb{P}(Z=1)=\alpha \in (0,1)$. Here $Z_i=1$ indicates that individual $i$ is in the treatment group and $Z_i=0$ indicates that $i$ is in the control group. We denote the high dimensional covariates as $X_i \sim X \in \mathbb{R}^p$. Under stable unit treatment value assumption (SUTVA \citep{rubin1978}), we denote the potential time to event outcome for individual $i$ with treatment $z\in \{0,1\}$ as $T_i^z$ and the corresponding potential censoring time as $C_i^z$ for $i=1,\cdots,n$. Also we use the consistency assumption \citep{cole2009} to link the observed and potential outcome, i.e., $T_i=Z_iT_i^1+(1-Z_i)T_i^0$ and $C_i=Z_iC_i^1+(1-Z_i)C_i^0$. Due to censoring, we cannot directly observe $(T_i,C_i)$, instead, we observe the composite outcome $(Y_i,\Delta_i)\in R^{+}\times \{0,1\}$, where $Y_i=T_i\wedge C_i=\min(T_i,C_i)$, and $\Delta_i=\mathbbm{1} (T_i\leq C_i)$, where $\mathbbm{1}(\cdot)$ is the indicator function. We observed the data $\mathcal{D}=\{\mathcal{D}_i : i=1,\cdots,n\}$ where $\mathcal{D}_i=(Y_i, \Delta_i, Z_i,X_i)$. We use the counting process notation with $N_i(t)=\mathbbm{1}(Y_i\leq t, \Delta_i=1)$ and $\mathcal{Y}_i(t)=\mathbbm{1}(Y_i\geq t)$ and we define the filtration $\mathcal{F}_t=(N_i(s),\mathcal{Y}_i(s),Z_i,X_i,s\in [0,t])$.

\subsection{Estimation}

For survival analysis, there are multiple choices of scales \citep{chen2001,hernan2010,vanderweele2011} to quantify ACE. Here we consider the ACE on the survival probability scale and define the finite sample ACE as
\begin{eqnarray}
\tau_{\text{fs}}(t)=n^{-1}\sum_{i=1}^n \left\{\mathbbm{1}(T_i^1>t)-\mathbbm{1}(T_i^0>t)\right\}.
\end{eqnarray}
When considering the finite sample is randomly sampled from a super population, we can define the super population ACE \citep{imbens2015} as
\begin{eqnarray}
\tau_{\text{sp}}(t)=\mathbb{E} \left[n^{-1}\sum_{i=1}^n \left\{\mathbbm{1}(T_i^1>t)-\mathbbm{1}(T_i^0>t)\right\}\right]=\mathbb{P}(T^1>t)-\mathbb{P}(T^0>t).
\end{eqnarray}


\paragraph{}
Under the randomization assumption, $\tau_{\text{sp}}(t)$ can be rewritten as 
\begin{equation*}
\tau_{\text{sp}}(t)=\mathbb{P}(T_i>t|Z_i=1)-\mathbb{P}(T_i>t|Z_i=0), 
\end{equation*}
which can be consistently estimated without adjustment for covariate $X$ if we observe all $T_i$'s, i.e., we can propose an estimator
\[ \tilde{\tau}_0(t) = n_1^{-1}\sum_{i:Z_i=1} \mathbbm{1}(T_i>t) -n_0^{-1}\sum_{i:Z_i=0} \mathbbm{1}(T_i>t). \]


We know that $\tilde{\tau}_0(t)$ is not useful because, due to the censoring in real data, we cannot observe all $T_i$'s. However, the IPCW technique can be used to form a crude unadjusted estimator: 
\begin{equation}\label{eqn:tau_0}
\hat{\tau}_{0}(t)=n_1^{-1}\sum_{i:Z_i=1}\frac{R_i(t) \mathbbm{1}(T_i>t)}{\hat{\pi}_i(t)}-n_0^{-1}\sum_{i:Z_i=0}\frac{R_i(t) \mathbbm{1}(T_i>t)}{\hat{\pi}_i(t)},
\end{equation}
where $n_0=\sum_{i=1}^n (1-Z_i)$  is the number of samples in the control group and $n_1=\sum_{i=1}^n Z_i$ is the number of samples in the treatment group. Here $R_i(t)=\mathbbm{1}(C_i>T_i\wedge t)$ and $\hat{\pi}_i(t)$ is any uniformly consistent estimator of $\pi_i(t)=\mathbb{P}(C>T\wedge t|T=T_i,Z=Z_i,X=X_i)$. For example, under random censoring assumption, we can estimate $\pi_i(t)$ using $\hat{\pi}_i(t)=\hat{S}_C(t\wedge T_i)$ where $\hat{S}_C(t)$ is the Kaplan Meier estimator \citep{kaplan1958} for $S_C(t)=\mathbb{P}(C>t)=\mathbb{P}(C>t|Z=z,X=x)$. When the censoring depends on $X$ or $Z$, we could use Kernel conditional Kaplan Meier estimator as in \citet{dabrowska1989} to estimate $S_C(t|z,x)=\mathbb{P}(C>t|Z=z,X=x)$, which is known to satisfy the uniform consistency requirement. 

When $T_i\leq t$, we have $R_i(t) \mathbbm{1}(T_i>t)=0$, so \eqref{eqn:tau_0} can be further simplified by

\begin{equation} \label{tau0}
\hat{\tau}_{0}(t)=n_1^{-1}\sum_{i:Z_i=1}\frac{R_i(t) \mathbbm{1}(T_i>t)}{\hat{S}_C(t|Z_i,X_i)}-n_0^{-1}\sum_{i:Z_i=0}\frac{R_i(t) \mathbbm{1}(T_i>t)}{\hat{S}_C(t|Z_i,X_i)}.
\end{equation}
As a remark, since in constructing $\hat{\tau}_{0}(t)$ we do not do any covariate adjustments, a lost in efficiency is expected. We will show this both theoretically (in Theorem \ref{thm:4}) and with numerical studies later.   

To utilize the covariate information, an alternative way to estimate $\tau_{\text{sp}}(t)$ is based on the representation: 
\begin{equation*}
\tau_{\text{sp}}(t)=\int \left\{\mathbb{P}(T>t|Z=1,X=x)-\mathbb{P}(T>t|Z=0,X=x)\right\}dF_{X}(x).
\end{equation*}
We can construct an estimator $\tilde{\tau}_1(t)$ as below:
\begin{equation}
\tilde{\tau}_1(t)=n^{-1}\sum_{i=1}^n \left\{\hat{\mu}^{(1)}(t,X_i)-\hat{\mu}^{(0)}(t,X_i)\right\},
\end{equation}
where $\hat{\mu}^{(z)}(t,X_i)$ is an estimator of \[\mu^{(z)}(t,X_i):= \mathbb{P}(T^z>t|X=x)=\mathbb{P}(T>t|Z=z,X=x).\] When nonparametric model is used to estimate $\mu^{(z)}(\cdot)$, usually, we consider the leave one out prediction $\hat{\mu}^{(z,-i)}(\cdot)$, and the corresponding estimator for $\tau_{\text{sp}}(t)$ can be written as:
\begin{equation} \label{tau1}
\hat{\tau}_1(t)=n^{-1}\sum_{i=1}^n \left\{\hat{\mu}^{(1,-i)}(t,X_i)-\hat{\mu}^{(0,-i)}(t,X_i)\right\}.
\end{equation}
Here $\hat{\mu}^{(z,-i)}(t,x)$ is a uniformly consistent estimator for $\mu^{(z)}(t,x)$ that does not depend on the $i$-th training data asymptotically. For example, we can construct the leave-one-out estimator using Cox model \citep{cox1972}, penalized survival regression model \citep{tibshirani1997} or survival random forest \citep{ishwaran2008,ishwaran2010}. For semi-parametric models such as Cox model, there is no tuning parameter to choose and including the $i$-th individual to fit the model is asymptotically equivalent to fitting without the $i$-th individual. However, for the high-dimensional regression models such as random forest, it is necessary to use the out of bag prediction to remove the strong correlation between $(Y_i,\Delta_i)$ and $\hat{\mu}^{(z)}(t,X_i)$.

We expect $\hat{\tau}_1(t)$ to have improved efficiency compared with $\hat{\tau}_0(t)$, as it uses the covariate information. However, the limiting distribution of $\hat{\tau}_1(t)$ is hard to characterize and might not be Gaussian, leading to challenges in studying the efficiency and making inference.  

Alternatively, suppose we observe all $T_i$'s, high-dimensional regression adjustments can be done to construct the following estimator:
{\small
\begin{eqnarray*}
\tilde{\tau}_{2}(t)&=&n^{-1}\sum_{i=1}^n\left\{\hat{\mu}^{(1,-i)}(t,X_i)-\hat{\mu}^{(0,-i)}(t,X_i)\right\}+n_1^{-1}\sum_{i:Z_i=1}\left\{\mathbbm{1}(T_i> t)-\hat{\mu}^{(1,-i)}(t,X_i)\right\}\\
&&-n_0^{-1}\sum_{i:Z_i=0}\left\{\mathbbm{1}(T_i> t)-\hat{\mu}^{(0,-i)}(t,X_i)\right\}.
\end{eqnarray*}}
This estimator can be viewed as an extension of the high-dimensional covariate adjustment for linear models proposed by \citet{wager2016}. The idea is that with the individuals whose outcomes are observed, we will use the observed outcome instead of their estimates in the ACE estimation. With survival data, we cannot observe all $T_i$'s because of censoring. Using IPCW technique \citep{robins2000}, we propose two model adjusted estimators based on observable data as below:
{\small
\begin{multline}\label{tau2}
\hat{\tau}_2(t)=n^{-1}\sum_{i=1}^n \left\{\hat{\mu}^{(1,-i)}(t,X_i)-\hat{\mu}^{(0,-i)}(t,X_i)\right\} \\+n_1^{-1}\sum_{i:Z_i=1} \left\{\frac{R_i(t)\mathbbm{1}(T_i>t)}{\hat{\pi}_i(t)}-\hat{\mu}^{(1,-i)}(t,X_i)\right\} \\
-n_0^{-1}\sum_{i:Z_i=0}\left\{\frac{R_i(t)\mathbbm{1}(T_i>t)}{\hat{\pi}_i(t)}-\hat{\mu}^{(0,-i)}(t,X_i)\right\},
\end{multline}}
and
{\small
\begin{eqnarray*}
\hat{\tau}_3(t)&=&n^{-1}\sum_{i=1}^n \left\{\hat{\mu}^{(1,-i)}(t,X_i)-\hat{\mu}^{(0,-i)}(t,X_i)\right\}+n_1^{-1}\sum_{i:Z_i=1} \frac{R_i(t)\left\{\mathbbm{1}(T_i>t)-\hat{\mu}^{(1,-i)}(t,X_i)\right\}}{\hat{\pi}_i(t)}
\end{eqnarray*}
\begin{eqnarray} \label{tau3}
&&-n_0^{-1}\sum_{i:Z_i=0}\frac{R_i(t)\left\{\mathbbm{1}(T_i>t)-\hat{\mu}^{(0,-i)}(t,X_i)\right\}}{\hat{\pi}_i(t)}.
\end{eqnarray}}

Here $\hat{\tau}_3(t)$ can be viewed as using model prediction to estimate the individual causal effect among those non-censored individuals and then use the inverse probability censoring weighting (IPCW) to obtain the adjustment needed for these groups. In the following two sections we will show that under mild model assumptions, $\hat{\tau}_2(t)$ and $\hat{\tau}_3(t)$ are uniformly consistent estimators with improved efficiency than $\hat{\tau}_0(t)$ both theoretically and with numerical studies.

\section{Main results}
\subsection{Regularity Conditions}
\label{sec1subsec2}
Before we illustrate the theoretical guarantees about the performance of the proposed estimators $\hat{\tau}_0(t)$, $\hat{\tau}_1(t)$, $\hat{\tau}_2(t)$, $\hat{\tau}_3(t)$, we state our regularization conditions beyond the well-known assumptions for causal inference (such as SUTVA and consistency stated in Section \ref{sec1subsec1}). For different estimators, we assume different conditions and the details are stated in the theorems. \\






\begin{condition} \label{ass:rand_cens}
\textbf{(Random censoring)}
$C^z$ is independent of $T^z$ conditioning on $(X,Z)$. Also, we assume that for the maximum time of interest $t_0$, there exists a positive number $\delta>0$ such that $\mathbb{P}(Y>t_0|Z=z,X=x)=\mathbb{P}(C>t_0|Z=z,X=x)\mathbb{P}(T>t_0|Z=z,X=x)\geq \delta$ for all $z,x$.
\end{condition}

Condition \ref{ass:rand_cens} is a commonly used condition in survival analysis.  

\begin{condition} \label{cond:1a} 
We assume that the estimator $\hat{S}_C(t|z,x)$ satisfy that \[\sup_{\substack{z\in \{0,1\},\\ x\in \mathbb{X},\\t\in [0,t_0]}}|\hat{S}_C(t|z,x)-S_C(t|z,x)| \stackrel{p}{\rightarrow} 0.\]
\end{condition}

This condition is the uniform consistency of $\hat{S}_C(t|z,x)$, this condition is automatically satisfied when nonparametric model, such as Kernel conditional Kaplan Meier, is used to obtain $\hat{S}_C(t|z,x)$. Condition \ref{cond:1a} is satisfied under condition \ref{cond:1c}.

\begin{condition} \label{cond:1c}  We assume that there exists an influence function $\psi(t,z,x;\mathcal{D})$ that is equicontinuous, with finite variance process (i.e., $\mathbb{E}[\psi(t,z,x;\mathcal{D})^2]\leq \infty$), satisfying 
\[\sqrt{n}(\hat{\pi}_i(t)-\pi_i(t))=n^{-1/2}\sum_{j=1}^n\psi(t,Z_i,X_i;\mathcal{D}_j)+\textnormal{Rem}(i,t),\]
where $\sup_{i,t\in[0,t_0]}|\textnormal{Rem}(i,t)|\stackrel{p}{\rightarrow} 0$, and that the influence function is separable. i.e. there exists a finite $K$, \[\psi(t,z,x;\mathcal{D}_j)=\sum_{k=1}^Kf_k(z,x)\psi_k(t;\mathcal{D}_j).\]
\end{condition}

This condition is the regular and asymptotic linearity of $\hat{\pi}_i(t)$ with separable influence function.

\begin{condition} \label{cond:2}  We assume that the estimator $\hat{\mu}^{(z,-i)}(t|x)$ is uniformly consistent, that 
\[\sup_{\substack{z\in \{0,1\},\\ x\in \mathbb{X},\\t\in [0,t_0]}}|\hat{\mu}^{(z,-i)}(t|x)-\mu^{(z)}(t|x)|\stackrel{p}{\rightarrow} 0.\] 
\end{condition}

Condition \ref{cond:2} states about the uniform consistency of $\hat{\mu}^{(z,-i)}(t|x)$; it is automatically satisfied under Condition \ref{cond:4}.


\begin{condition} \label{cond:4} \textbf{(Jackknife compatibility)} We assume that $\hat{\mu}^{(z)}(t, X)$ is jackknife compatible; which is to say the expected jackknife estimator \citep{efron1981} of the variance process for $\hat{\mu}^{(z)}(t,X)$ converges to 0 uniformly over $t\in [0,t_0]$ and $X\in \mathbb{X}$; i.e. for all $z\in \{0,1\}$, $t\in [0,t_0]$ and $X \in \mathbb{X}$, we have
\begin{eqnarray*}
\mathbb{E}\left[\sum_{i:Z_i=z}\left\{\hat{\mu}^{(z,-i)}(t, X)-\hat{\mu}^{(z)}(t, X)\right\}^2|n_z\right]\leq a(n_z)
\end{eqnarray*}
with some sequence $a(n_z)\rightarrow 0$ which does not depend on $t$ and $X$.
\end{condition}

The jackknife compatibility implies that the jackknife estimate of variance for $\hat{\mu}^{(z)}$ converges to zero in expectation. This is a weak condition that many classical estimators satisfy this condition. \cite{wager2016} showed that Random Forest (RF) estimator also satisfy this condition.

\begin{condition} \label{cond:5}\textbf{(Risk consistency)} We assume $\hat{\mu}^{(z)}(t,X)$ is `risk-consistent' to $\mu^{(z)}(t,X)$. i.e. 
\begin{eqnarray*}
\mathbb{E}\left[\sum_{i:Z_i=z}\left\{\hat{\mu}^{(z)}(t,X_i)-\mu^{(z)}(t,X_i)\right\}^2|n_z\right]\leq a(n_z)
\end{eqnarray*}
uniformly over $t \in [0,t_0]$, for $z\in \{0,1\}$, $X\in \mathbb{X}$, with some $a(n_z)\rightarrow 0$ with $n_z\rightarrow \infty$.\\
\end{condition}

The risk consistency condition is mild that it is satisfied by most estimators. In particular, Under the assumption of random censoring, \citet{ishwaran2008} showed that survival RF estimator satisfies this assumption.

\subsection{Asymptotics}
\label{sec1subsec3}
In this section, we first show the consistency result of the proposed estimators $\hat{\tau}_0, \hat{\tau}_1, \hat{\tau}_2, \hat{\tau}_3$ (Theorem \ref{thm:1}). Then we show the asymptotic normality results for the estimators (Theorems \ref{thm:2} and \ref{thm:3}). Finally we compare the efficiency of the proposed estimators (Theorem \ref{thm:4}). 

\paragraph{}
\begin{theorem} \label{thm:1}  Under Conditions \ref{ass:rand_cens} and \ref{cond:1a}, the crude estimator $\hat{\tau}_0$ is uniformly consistent. Under Condition \ref{cond:2}, the model-based estimator $\hat{\tau}_1$ is uniformly consistent. Under Conditions \ref{ass:rand_cens}, \ref{cond:1a}, and \ref{cond:2}, the high dimensional adjustment estimators $\hat{\tau}_2$, $\hat{\tau}_3$ are also uniformly consistent estimators of $\tau_{\text{sp}}(t)$ for all $t\in [0,t_0]$. 
\end{theorem} 

The proof of Theorem \ref{thm:1} is postponed to Appendix \ref{app:proof}. Under a stronger assumption of completely random censoring, we can simply use Kaplan Meier to estimate $\hat{S}_C(t)=\hat{S}_C(t|z,x)$. Alternatively, for semiparametric models such as Cox, we need the model to be correctly specified for condition \ref{cond:1a} to hold. Condition \ref{cond:2} is automatically satisfied when a nonparametric model, such as survival random forest, is used to obtain the estimator $\hat{\mu}^{(z,-i)}(t|x)$. For semiparametric models such as Cox regression, we need the model to be correctly specified.

\paragraph{}

The limiting distribution of $\hat{\tau}_1$ is in general hard to obtain and might not be Gaussian. For example, when $L_1$-penalized estimator is used for constructing $\hat{\mu}^{(z)}$, \citet{knight2000} showed that the limiting distribution is not tractable and can have mass at 0. So we will not discuss the asymptotics results for $\hat{\tau}_1$. Although $\hat{\mu}^{(z)}$ also appears in the estimators $\hat{\tau}_2$ and $\hat{\tau}_3$, its bias can be canceled based on the randomization of $Z$. To further obtain the asymptotic normality for $\hat{\tau}_2$ and $\hat{\tau}_3$, we need additional assumptions. Here we summarize the asymptotics for $\hat{\tau}_0$, $\hat{\tau}_2$ and $\hat{\tau}_3$ in the following two theorems. The proofs can be found in the Appendix \ref{app:proof}.
\paragraph{}
\begin{theorem} \label{thm:2} 
Under Conditions \ref{ass:rand_cens} and \ref{cond:1c}, the proposed estimator $\hat{\tau}_0(t)$ satisfies that $\hat{U}_0(t) :=\sqrt{n}\left\{\hat{\tau}_0(t)-\tau_{\text{sp}}(t)\right\}$ weakly converges to a Gaussian process $U_0(t)$ with mean $0$ and covariance process $\Sigma_0(t,s)=Cov(U_0(t),U_0(s))$, where the form and a consistent estimation can be found in the Appendix \ref{app:proof}.
\end{theorem} 
\paragraph{}
\begin{theorem} \label{thm:3} 
Denote \[\hat{U}_2(t):=\sqrt{n}\left\{\hat{\tau}_2(t)-\tau_{\text{sp}}(t)\right\} \quad \textnormal{and} \quad \hat{U}_3(t):=\sqrt{n}\left\{\hat{\tau}_3(t)-\tau_{\text{sp}}(t)\right\}.\] Under Conditions \ref{ass:rand_cens}, \ref{cond:1c}, \ref{cond:4}, and \ref{cond:5}, $\hat{U}_2(t)$ and $\hat{U}_3(t)$ weakly converge to Gaussian processes $U_2(t)$ and $U_3(t)$ with mean $0$ and covariance processes $\Sigma_2(t,s)=Cov(U_2(t),U_2(s))$ and $\Sigma_3(t,s)=Cov(U_3(t),U_3(s))$, where the form and a consistent estimation can be found in the Appendix \ref{app:proof}. 
\end{theorem} 
\paragraph{}
In practice, for $\hat{\tau}_2$ and $\hat{\tau}_3$, the asymptotic variance from the IPCW is often smaller than the variation in $\mathbbm{1}(T_i>t)$. Therefore we have the following results regarding the efficiency comparing the three estimators $\hat{\tau}_0$, $\hat{\tau}_2$ and $\hat{\tau}_3$. 
\paragraph{}
\begin{theorem} \label{thm:4} Under Conditions \ref{ass:rand_cens},\ref{cond:1c}, \ref{cond:4}, and \ref{cond:5}, and $S_C(t|z,x)=S_C(t)$, the estimators $\hat{\tau}_2(t)$, $\hat{\tau}_3(t)$ are asymptotic more efficient than $\hat{\tau}_0(t)$ in the sense that for any fixed non-negative weight function $w(t)$, we have \[Var\left\{\int _0^{t_0}w(t)\hat{\tau}_3(t)dt\right\}\leq Var\left\{\int _0^{t_0}w(t)\hat{\tau}_2(t)dt\right\}\leq Var\left\{\int_0^{t_0} w(t)\hat{\tau}_0(t)dt\right\}\] holds asymptotically. The second equal sign holds only when $X$ has no effect on $T$, i.e., $\mu^{(z)}(t,X_i)=\mu^{(z)}(t)$ with probability 1 for all $t$ where $w(t)$ is non-zero. \end{theorem} 
The proof details are postponed in Appendix \ref{app:proof}.

\section{Simulation}
\label{sec:2}

\begin{figure}
\includegraphics[width=4.5in]{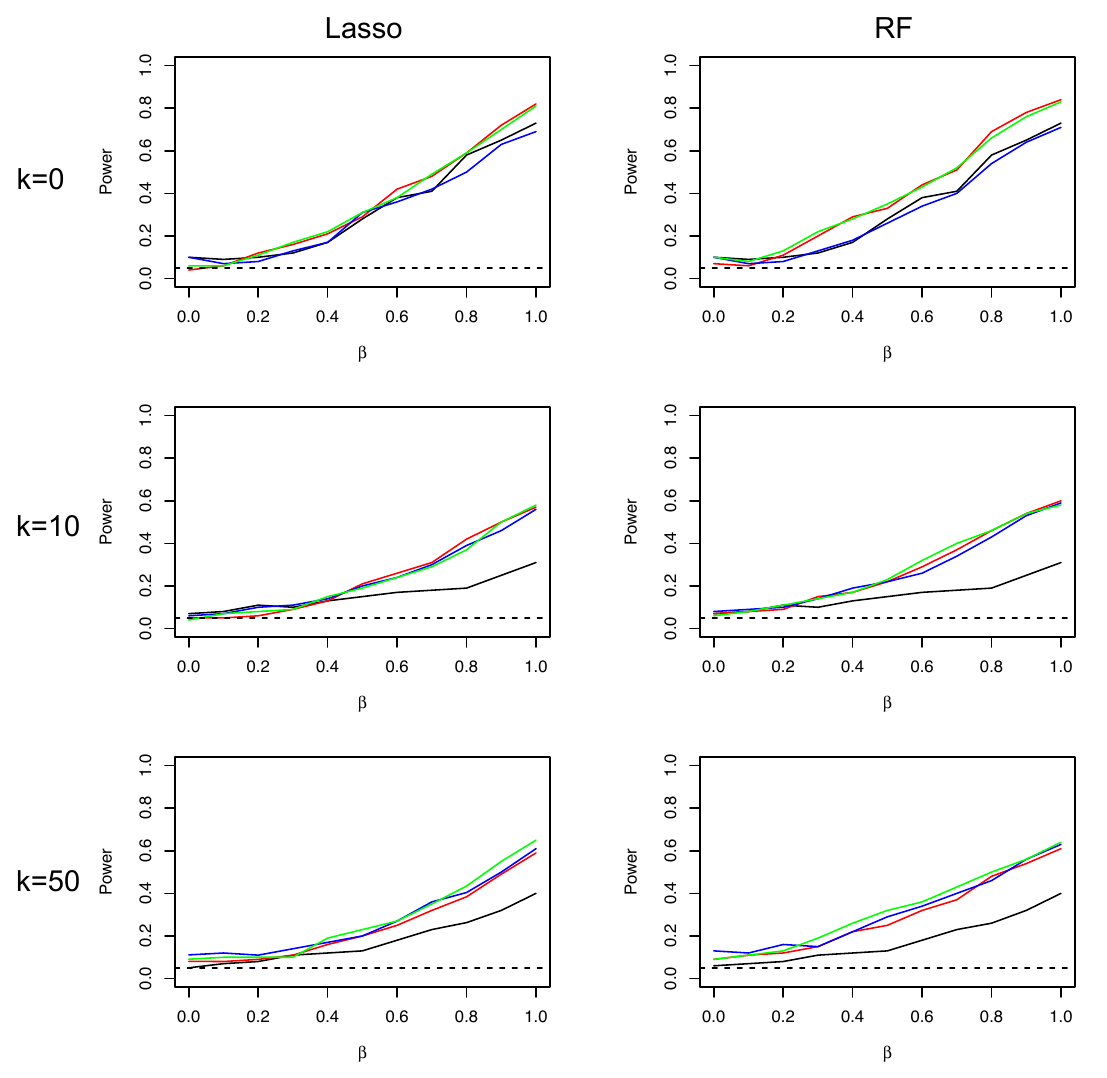}
\caption{Power curve for different estimators under high dimensional setting ($p=50$) when $s_0=s_1=1$. Different estimators are presented with different colors as below: $\hat{\tau}_0$, black; $\hat{\tau}_1$, red; $\hat{\tau}_2$, blue; $\hat{\tau}_3$, green.}
\label{fig1}
\end{figure}

\paragraph{Settings}
We study the performance of our proposed estimators on data with different dimensionality and sparsity levels. We let $n=100$, $\alpha=0.5$ and the number of variables ($p$) and true sparsity ($k$) as $(p,k)=(10,10)$, $(50,10)$ or $(50,50)$. We generate $X$ from multivariate normal distribution with $AR(1)$ correlation structure, and the autocorrelation parameter $\rho$ is set at $\rho=0.8$. Survival time $T^0$ and $T^1$ are independently sampled using Cox model with hazard functions
\[\lambda_0(t)=\exp(X\gamma_0) \quad \text{and} \quad \lambda_1(t)=\exp(\beta+X\gamma_1), \]
where $\gamma_{0j}=\frac{s_0\mathbbm{1}(j\leq k)}{j}$ and $\gamma_{1j}=\frac{s_1\mathbbm{1}(j\leq k)}{j}$ for $j=1,\cdots,p$ are parameters indicating the covariate effect and $s_0$, $s_1$ represent the overall effect magnitude. We use censoring distribution $C\sim \text{Unif}[0,2.5]$ to obtain $50\%-70\%$ event rate. We set $t_0$ of interest as the median observation time calculated from a large sample (N=50,000) of observed $Y$ pooling treatment and control groups. We vary $\beta$ from 0 to 1, and let $s_0=s_1=s$ or $s_0=0$ and $s_1=s$ with $s$ changing from 0 to 1.  For each data setting we perform 100 simulations to study the performance of the four estimators, $\hat{\tau}_0(t_0)$, $\hat{\tau}_1(t_0)$, $\hat{\tau}_2(t_0)$, $\hat{\tau}_3(t_0)$ defined in (\ref{tau0}), (\ref{tau1}), (\ref{tau2}) and (\ref{tau3}) paired with three different models (Cox, Lasso and Random Forest). Since for some estimators, the asymptotic distribution is intractable, we use Bootstrap to construct the nominal 95\% confidence intervals and evaluate their empirical coverage rate (CR) and power (proportion rejected at significance level 0.05) in finite sample cases.
\paragraph{Results}
We study the performance of our proposed estimators in both low dimensional and high dimensional cases. Simulation results for low-dimensional settings are postponed in Appendix B. For high-dimensional settings, we only focus on Lasso type adjustment and Random Forest adjustment, because Cox adjustment is not applicable. In Figure \ref{fig1}, we show the power curve of the four estimators from two different models (Lasso and Random Forest) when there is no $X$ effect ($k=0$) and there is $X$ effect ($s_0=s_1=1$) under high dimensional settings with sparsity, i.e., $(p,k)=(50,10)$ and high dimensional setting without sparsity, i.e.,$(p,k)=(50,50)$. From Figure \ref{fig1}, the performance of Random Forest and Lasso are similar; $\hat{\tau}_3$ outperforms $\hat{\tau}_0$ and $\hat{\tau}_2$ no matter the covariate effect exists or not. When there is no covariate effect, $\hat{\tau}_2$ has some slight power loss, but $\hat{\tau}_2$ is better than $\hat{\tau}_0$ when covariate effects exist. $\hat{\tau}_2$'s power lies in between $\hat{\tau}_0$ and $\hat{\tau}_3$, which is consistent with our theoretical findings.

\begin{figure}
\includegraphics[width=4.5in]{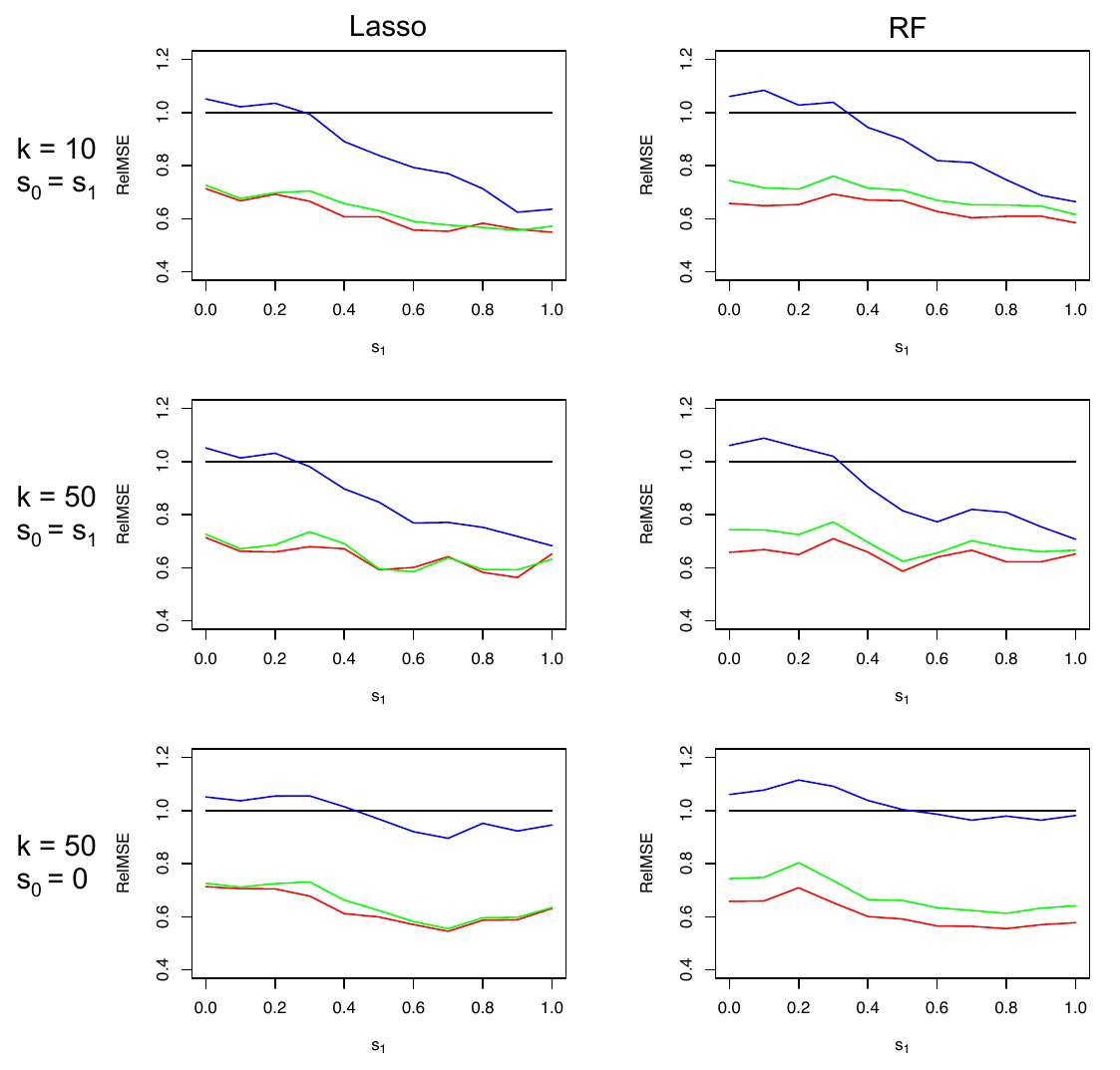}
\caption{Relative efficiency with the change of covariate effect for different estimators under high dimensional setting ($p=50$) when $\beta=0.5$. Different estimators are presented with different colors as below: $\hat{\tau}_0$, black; $\hat{\tau}_1$, red; $\hat{\tau}_2$, blue; $\hat{\tau}_3$, green.}
\label{fig2}
\end{figure}

\paragraph{}
In Figure \ref{fig2}, we show the relative MSE of the four estimators, $\hat{\tau}_0$, $\hat{\tau}_1$, $\hat{\tau}_2$, $\hat{\tau}_3$ with respect to the change of effect size of $X$ ($s_1=0,0.1,\cdots,1$) when there is no interaction (left: $s_0=s_1$), or there are interactions (right: $s_0=0$, $s_1=s$) under high-dimensional settings ($p=50$) for $\beta=0.5$. From Figure \ref{fig2}, we see that $\hat{\tau}_0$ and $\hat{\tau}_2$ are both less efficient than $\hat{\tau}_3$. The relative MSEs for the adjustment methods $\hat{\tau}_1$, $\hat{\tau}_2$ and $\hat{\tau}_3$ all decrease when the covariate effect increases. Similar results hold for other $\beta$'s.\\
\paragraph{}
In Appendix B, we show additional simulation results for the comparison of Cox adjustment, Lasso adjustment and Random Forest adjustment in low dimensional settings. In Figure S1, we show the power change of the four estimators paired with the three different models with respect to the change of effect size $\beta$ with or without the covariate effect under the low dimensional setting where $(p,k)=(10,10)$. We see that the overall performance of the three different models are about the same; for all of them, $\hat{\tau}_1$ and $\hat{\tau}_3$ perform better than $\hat{\tau}_0$ no matter the covariate effect exists or not. When there is no covariate effect, $\hat{\tau}_2$ has some slight power loss, but $\hat{\tau}_2$ is better than $\hat{\tau}_0$ when covariate effects exist. 

\paragraph{}
In Figure S2, we show the relative MSE of the four estimators from three different models respective to the change of covariate effect ($s_1=0,0.1,\cdots,1$) with interaction ($s_1=s, s_0=0$) or without interaction ($s_0=s_1=s$) under low dimensional setting, $(p,k)=(10,10)$ with $\beta=0.5$. From figure S2, we see similar results comparing to the high dimensional settings. Similar results hold for other $\beta$'s.  

\paragraph{}
For some representative settings, we report the bias, SD, mean estimated standard error (ESE), and CR for these estimators in Table 1. From the table, we can see that all estimators are unbiased and with CRs close to $95\%$, the ESE's are close to SD's under all scenarios. The Cox model usually does not converge when number of covariates are large, so we did not present its result. 

\begin{table}
\caption{Simulation Results}
\label{table1}
\begin{tabular}{|c|c|c|ccccc|}
\hline
Method &($\beta$,$p$,$k$,$s_0$,$s_1$) &Estimator &Bias &SD & ESE &RelMSE &CR \\
\hline
\multirow{36}{*}{Lasso}&\multirow{4}{*}{$(0,10,10,0,0)$} & $\hat{\tau}_0(t)$ &0.032&0.132&0.126&1.000&0.90\\
&& $\hat{\tau}_1(t)$ &0.018&0.102&0.108&0.581&0.92\\
&& $\hat{\tau}_2(t)$ &0.035&0.132&0.129&1.014&0.91\\
&& $\hat{\tau}_3(t)$ &0.022&0.105&0.109&0.622&0.92\\
\cline{2-8}
&\multirow{4}{*}{$(0.5,10,10 ,0 ,0)$} & $\hat{\tau}_0(t)$ &0.014&0.126&0.118&1.000&0.94\\
&& $\hat{\tau}_1(t)$ &0.007&0.104&0.106&0.675&0.94\\
&& $\hat{\tau}_2(t)$ &0.017&0.127&0.120&1.024&0.93\\
&& $\hat{\tau}_3(t)$ &0.009&0.107&0.106&0.715&0.93\\
\cline{2-8}
&\multirow{4}{*}{$(0.5,10,10 ,0.5 ,0.5)$} & $\hat{\tau}_0(t)$ &0.007&0.109&0.111&1.000&0.95\\
&& $\hat{\tau}_1(t)$ &0.003&0.084&0.087&0.586&0.94\\
&& $\hat{\tau}_2(t)$ &0.006&0.099&0.094&0.822&0.93\\
&& $\hat{\tau}_3(t)$ &0.003&0.086&0.087&0.616&0.92\\
\cline{2-8}
&\multirow{4}{*}{$(0,50,10 ,0 ,0)$} & $\hat{\tau}_0(t)$ &0.032&0.131&0.126&1.000&0.90\\
&& $\hat{\tau}_1(t)$ &0.013&0.102&0.110&0.589&0.96\\
&& $\hat{\tau}_2(t)$ &0.032&0.133&0.129&1.039&0.90\\
&& $\hat{\tau}_3(t)$ &0.017&0.104&0.110&0.617&0.94\\
\cline{2-8}
&\multirow{4}{*}{$(0.5,50,10 ,0 ,0)$} & $\hat{\tau}_0(t)$&0.013&0.125&0.118&1.000&0.94\\
&& $\hat{\tau}_1(t)$ &0.002&0.106&0.106&0.714&0.95\\
&& $\hat{\tau}_2(t)$ &0.013&0.128&0.120&1.052&0.94\\
&& $\hat{\tau}_3(t)$ &0.004&0.107&0.106&0.727&0.96\\
\cline{2-8}
&\multirow{4}{*}{$(0.5,50,10,0.5,0.5)$} & $\hat{\tau}_0(t)$&0.008&0.109&0.111&1.000&0.94\\
&& $\hat{\tau}_1(t)$ &0.003&0.086&0.091&0.608&0.94\\
&& $\hat{\tau}_2(t)$ &0.007&0.100&0.097&0.839&0.94\\
&& $\hat{\tau}_3(t)$ &0.003&0.087&0.089&0.631&0.94\\
\cline{2-8}
&\multirow{4}{*}{$(0,50,50,0,0)$} & $\hat{\tau}_0(t)$ &0.032&0.131&0.126&1.000&0.90\\
&& $\hat{\tau}_1(t)$ &0.013&0.102&0.110&0.589&0.96\\
&& $\hat{\tau}_2(t)$ &0.032&0.133&0.129&1.039&0.90\\
&& $\hat{\tau}_3(t)$ &0.017&0.104&0.110&0.617&0.94\\
\cline{2-8}
&\multirow{4}{*}{$(0.5,50,50, 0,0)$} & $\hat{\tau}_0(t)$ & 0.013&0.125&0.118&1.000&0.94\\
&& $\hat{\tau}_1(t)$ &0.002&0.106&0.106&0.714&0.95\\
&& $\hat{\tau}_2(t)$ &0.014&0.128&0.120&1.052&0.94\\
&& $\hat{\tau}_3(t)$ &0.004&0.107&0.106&0.727&0.96\\
\cline{2-8}
&\multirow{4}{*}{$(0.5,50,50,0.5,0.5)$} & $\hat{\tau}_0(t)$ & 0.005&0.110&0.111&1.000&0.94\\
&& $\hat{\tau}_1(t)$ &0.002&0.084&0.097&0.593&0.95\\
&& $\hat{\tau}_2(t)$ &0.004&0.101&0.108&0.848&0.90\\
&& $\hat{\tau}_3(t)$ &0.003&0.085&0.096&0.596&0.93\\
\hline
\multirow{36}{*}{RF} &\multirow{4}{*}{$(0,10,10,0,0)$} & $\hat{\tau}_0(t)$ &0.031&0.131&0.126&1.000&0.90\\
&& $\hat{\tau}_1(t)$ &0.014&0.100&0.100&0.560&0.94\\
&& $\hat{\tau}_2(t)$ &0.035&0.134&0.125&1.054&0.90\\
&& $\hat{\tau}_3(t)$ &0.017&0.107&0.101&0.651&0.91\\
\cline{2-8}
&\multirow{4}{*}{$(0.5,10,10 ,0 ,0)$} & $\hat{\tau}_0(t)$ &0.013&0.125&0.118&1.000&0.94\\
&& $\hat{\tau}_1(t)$ &0.004&0.103&0.098&0.680&0.95\\
&& $\hat{\tau}_2(t)$ &0.013&0.129&0.116&1.075&0.95\\
&& $\hat{\tau}_3(t)$ &0.002&0.109&0.099&0.761&0.95\\
\cline{2-8}
&\multirow{4}{*}{$(0.5,10,10 ,0.5 ,0.5)$} & $\hat{\tau}_0(t)$ &0.007&0.109&0.111&1.000&0.95\\
&& $\hat{\tau}_1(t)$ &-0.002&0.089&0.086&0.651&0.91\\
&& $\hat{\tau}_2(t)$ &0.002&0.104&0.094&0.907&0.93\\
&& $\hat{\tau}_3(t)$ &-0.002&0.092&0.085&0.701&0.90\\
\cline{2-8}
&\multirow{4}{*}{$(0,50,10 ,0 ,0)$} & $\hat{\tau}_0(t)$ &0.032&0.131&0.126&1.000&0.90\\
&& $\hat{\tau}_1(t)$ &0.013&0.100&0.099&0.558&0.93\\
&& $\hat{\tau}_2(t)$ &0.034&0.133&0.125&1.047&0.90\\
&& $\hat{\tau}_3(t)$ &0.016&0.107&0.100&0.647&0.90\\
\cline{2-8}
&\multirow{4}{*}{$(0.5,50,10 ,0 ,0)$} & $\hat{\tau}_0(t)$ &0.013&0.125&0.118&1.000&0.94\\
&& $\hat{\tau}_1(t)$ &0.004&0.102&0.098&0.658&0.96\\
&& $\hat{\tau}_2(t)$ &0.013&0.129&0.116&1.061&0.93\\
&& $\hat{\tau}_3(t)$ &0.002&0.108&0.098&0.744&0.96\\
\cline{2-8}
&\multirow{4}{*}{$(0.5,50,10,0.5,0.5)$} & $\hat{\tau}_0(t)$ &0.008&0.109&0.111&1.000&0.94\\
&& $\hat{\tau}_1(t)$ &0.001&0.090&0.087&0.668&0.93\\
&& $\hat{\tau}_2(t)$ &0.004&0.104&0.095&0.899&0.92\\
&& $\hat{\tau}_3(t)$ &0.000&0.092&0.086&0.708&0.91\\
\cline{2-8}
&\multirow{4}{*}{$(0,50,50,0,0)$} & $\hat{\tau}_0(t)$ &0.032&0.133&0.126&1.000&0.9\\
&& $\hat{\tau}_1(t)$ &0.013&0.100&0.099&0.558&0.93\\
&& $\hat{\tau}_2(t)$ &0.034 &0.133 &0.125&1.047 &0.90\\
&& $\hat{\tau}_3(t)$ &0.016&0.107&0.100&0.647&0.90\\
\cline{2-8}
&\multirow{4}{*}{$(0.5,50,50, 0,0)$} & $\hat{\tau}_0(t)$ &0.013&0.125&0.118&1.000&0.94\\
&& $\hat{\tau}_1(t)$ &0.004&0.102&0.098&0.658&0.96\\
&& $\hat{\tau}_2(t)$ &0.013&0.129&0.116&1.061&0.93\\
&& $\hat{\tau}_3(t)$ &0.002&0.108&0.098&0.744&0.96\\
\cline{2-8}
&\multirow{4}{*}{$(0.5,50,50,0.5,0.5)$} & $\hat{\tau}_0(t)$ &0.005&0.110&0.111&1.000&0.94\\
&& $\hat{\tau}_1(t)$ &0.003&0.084&0.093&0.587&0.93\\
&& $\hat{\tau}_2(t)$ &0.004&0.099&0.106&0.815&0.93\\
&& $\hat{\tau}_3(t)$ &0.001&0.087&0.092&0.625&0.91\\
\hline
\end{tabular}
\end{table}

\paragraph{}
In conclusion, with low dimensional covariate information, the adjustment by regression without variable selection improves MSE, while when the number of covariates increases, the Cox model does not converge and therefore variable selection methods like Lasso and Random Forest need to be considered for the adjustment. The proposed estimator $\hat{\tau}_3$ with random forest adjustment is shown to outperform in all scenarios when compared with $\hat{\tau}_0$ and $\hat{\tau}_2$ in terms of both power and relative efficiency, which is consistent with our theoretical results. Also, $\hat{\tau}_3$ has similar performance as $\hat{\tau}_1$ in our experimented data settings. Given the availability of its asymptotic properties, we recommend the use of $\hat{\tau}_3$ with Random Forest.

\section{Real data example}
\label{sec:3}
\begin{figure}
\includegraphics[width=4.5in]{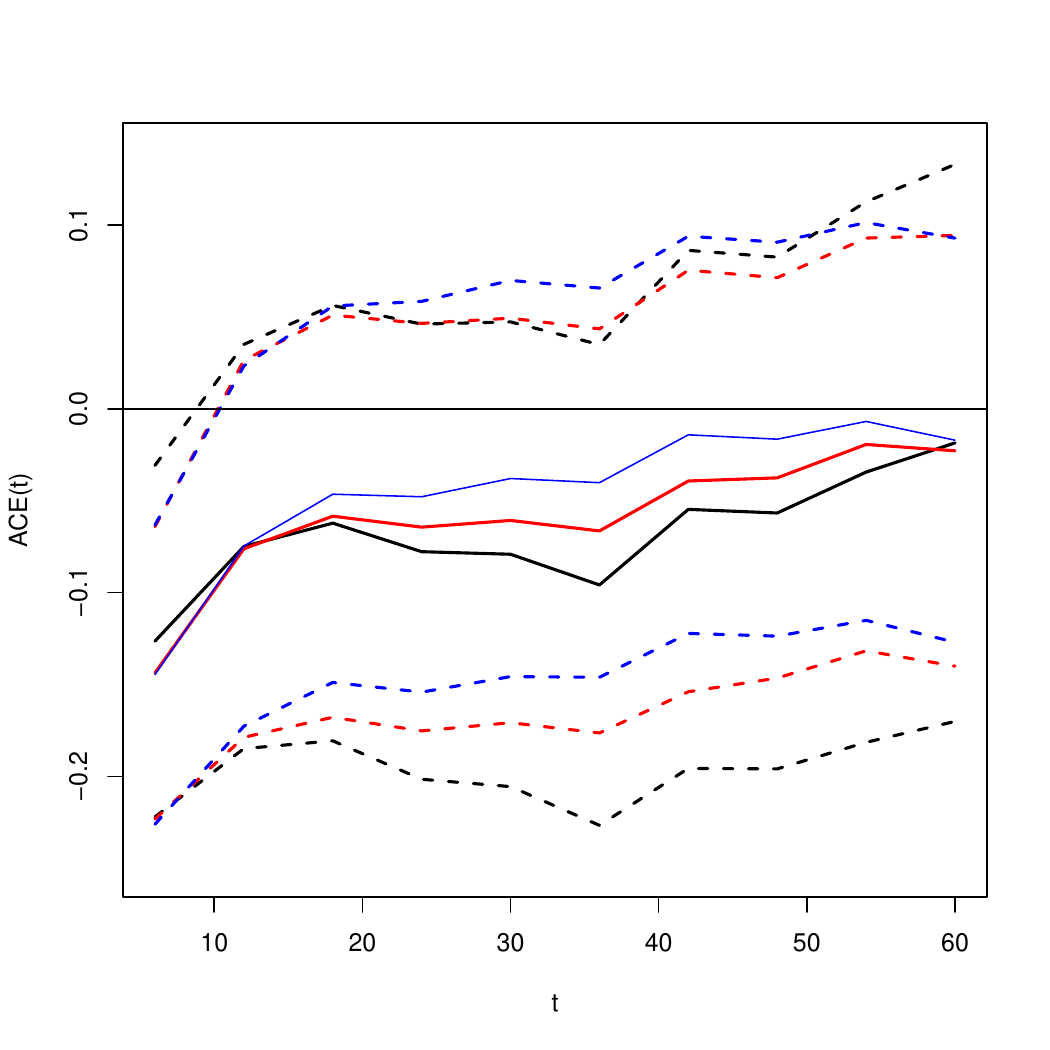}
\caption{Comparison of estimated average causal effect at time $t$ using crude estimator (black) and random forest adjusted estimators with short (red) and long (blue) list of covariate for adjustment. The dotted lines are point-wise 95\% confidence interval.}
\label{fig3}
\end{figure}

We apply our method to a real data example. Our data is from the Center for International Blood and Marrow Transplant Research (CIBMTR), on high risk Philadelphia-negative acute lymphoblastic leukemia (ALL) patients age 16 or older who underwent allogeneic hematopoietic stem cell transplantation (allo-HCT) in first complete remission (CR1) or second complete remission (CR2) between 1995 and 2011. The CIBMTR is comprised of clinical and basic scientists who share data on their blood and bone marrow transplant patients, with the CIBMTR Data Collection Center located at the Medical College of Wisconsin. The CIBMTR has a repository of information regarding the results of transplants at more than 450 transplant centers worldwide. Allo-HCT is a potential life-saving therapy for high risk ALL patients. We compare the overall survival probability between human leukocyte antigen (HLA) identical sibling donor (SIB) and 8/8 HLA-matched unrelated donor (MURD). It is impractical or impossible to conduct a randomized clinical trial for such comparison since SIB donors are available for only 20-30\% of all eligible patients. For the illustrative purpose, we use balancing propensity score approach to mimic a randomized trial. The data set used for our example with compete information consists of 523 SIB patients and 210 MURD patients, respectively. The variables considered in propensity score modeling and mimicking-trial-matching include cytogenetic abnormality (\textit{cytoabnorm}), conditioning regimen (\textit{condtbi}), Karnofsky score (\textit{kps}), graft-verse-host disease (GVHD) prophylaxis (\textit{gvhdgpc}), white blood count (\textit{wbcdxgp}), graft-type (\textit{graftype}), patient age (\textit{age}) and year of transplantation (\textit{yeargp}). Our pseudo-randomized trial cohorts consists of 156 SIB patients and 156 MURD patients, and all above adjusted variables are fully balanced between cohorts. We leave specific cytogenetic risk categories unmatched due to rare presence incidences (see Table S1 in Appendix C for detailed cytogenetic abnormality risk category list). The main purpose of this example is to show for the efficiency improvement using proposed estimator with adjustment of high dimensional covariates. Our main predictor of interest is \textit{MURD}. The short covariate list include \textit{condtbi}, \textit{yeargp}, \textit{del1q}, \textit{trisx}, \textit{t411}. The medium list further include \textit{kps}, \textit{age}, \textit{graftype}, \textit{gvhdgpc}, \textit{wbcdxgp}. The long list further include information on \textit{add5qdel5q}, \textit{add12pdel12p}, \textit{del7qm7}, \textit{m17i17qdel17p}, \textit{add7pi7q}, \textit{tratriphyper}, \textit{lohyperpo}, \textit{add9pdel9p}, \textit{t119}, \textit{t1011}, \textit{t1119}, \textit{del6q}, \textit{del11q}, \textit{tris8}, \textit{complex}, \textit{lohyper}, \textit{hihyper}, \textit{tetra}, \textit{mk}, \textit{axcomplex}, \textit{m13}, \textit{tris6}, \textit{tris10}, \textit{tris22}, \textit{add14q32}, \textit{del1q}, \textit{del17p}.
\paragraph{}
For each list type, we compared the four estimators under Cox adjustment, Lasso adjustment and Random Forest adjustment. We selected the time points from $t=6$ to $t=60$ month since transplantation and the estimated $ACE(t)$ and corresponding point-wise 95\% Confidence Interval for crude estimator $\hat{\tau}_0(t)$ and random-forest adjusted estimator $\hat{\tau}_3(t)$ with short and long covariate list for adjustment are shown in Figure \ref{fig3}. We also computed the average $ACE$ among the period and the results are shown in Table 2. From the results, we can see that the adjusted analysis provide us narrower confidence interval for both low and high-dimensional setting when comparing to the crude estimator.  

\begin{table}
\tiny
\caption{Estimates of average $ACE(t)$ with standard error (SE) and p-value among time 6 to 60 on real data}
\label{table2}
\begin{tabular}{|c|ccc|ccc|ccc|}
\hline
&\multicolumn{3}{|c|}{Cox Adjustment}&\multicolumn{3}{|c|}{Lasso Adjustment}&\multicolumn{3}{|c|}{Random Forest Adjustment}\\
\hline
\multicolumn{10}{|c|}{Short List Adjustment}\\
\hline
Estimator &Est &SE &p-value &Est &SE &p-value &Est &SE &p-value\\
\hline
$\hat{\tau}_0$ &-0.068 &0.059 &0.25 &-0.068 &0.059 &0.25 &-0.068 &0.059 &0.25\\
$\hat{\tau}_1$ &-0.071 &0.040 &0.08 &-0.079 &0.049 &0.11 &-0.064 &0.049 &0.19\\
$\hat{\tau}_2$ &-0.069 &0.046 &0.13 &-0.075 &0.056 &0.18 &-0.066 &0.057 &0.24\\
$\hat{\tau}_3$ &-0.059 &0.039 &0.13 &-0.074 &0.049 &0.13 &-0.059 &0.050 &0.24\\
\hline
\multicolumn{10}{|c|}{Medium List Adjustment}\\
\hline
Estimator &Est &SE &p-value &Est &SE &p-value &Est &SE &p-value\\
\hline
$\hat{\tau}_0$ &-0.068 &0.059 &0.25 &-0.068 &0.059 &0.25 &-0.068 &0.059 &0.25\\
$\hat{\tau}_1$ &-0.048 &0.052 &0.35 &-0.064 &0.055 &0.24 &-0.061 &0.050 &0.22\\
$\hat{\tau}_2$ &-0.055 &0.053 &0.30 &-0.068 &0.060 &0.26 &-0.059 &0.056 &0.29\\
$\hat{\tau}_3$ &-0.038 &0.051 &0.46 &-0.068 &0.055 &0.22 &-0.053 &0.050 &0.29\\
\hline
\multicolumn{10}{|c|}{Long List Adjustment}\\
\hline
Estimator &Est &SE &p-value &Est &SE &p-value &Est &SE &p-value\\
\hline
$\hat{\tau}_0$ &-0.068 &0.059 &0.25 &-0.068 &0.059 &0.25 &-0.068 &0.059 &0.25\\
$\hat{\tau}_1$ &NA &NA &NA &-0.064 &0.057 &0.26 &-0.052 &0.049 &0.29\\
$\hat{\tau}_2$ &NA &NA &NA &-0.068 &0.054 &0.21 &-0.050 &0.050 &0.31\\
$\hat{\tau}_3$ &NA &NA &NA &-0.068 &0.056 &0.22 &-0.045 &0.049 &0.37\\
\hline
\end{tabular}
\end{table}

\section{Discussion}
\label{sec:4}
In this work, we studied how to effectively use high-dimensional covariate information to reduce the variance of estimation for ACE under randomization trial. Both simulation studies and the real data example illustrate such efficiency gain compared with the crude IPCW estimator. The proposed estimators does not depend on either the semiparametric model (e.g., Cox) for the survival outcome or the assumption of homogeneity effect among population.
\paragraph{}
One strong assumption we made is random censoring. This assumption ensures that (1) the model for inverse probability censoring part is correctly specified and (2) the random forest estimator satisfy the risk consistency requirement. In general, as long as we can find a high-dimensional estimator that satisfies the risk consistency requirement, we can use a semiparametric model to estimate the censoring probability and extend our results to dependent censoring.
\paragraph{}
Also we would like to point out the augmentation part for our proposed estimator $\hat{\tau}_3$ might not be optimal. The use of augmented term similar to \citet{tsiatis2006} and \citet{lok2018} could be considered to increase efficiency.
Given that our simple estimator $\hat{\tau}_3$ already shows decent improvement in relative efficiency, here we choose not to use augmented IPCW with its optimal form in this work. 


\bibliographystyle{abbrvnat}      
\bibliography{refs}   

\appendix
\section{Additional proofs}\label{app:proof}

\subsection*{Proof of Theorem 1:}
We show the consistency of $\hat{\tau}_0(t)$, $\hat{\tau}_1(t)$, $\hat{\tau}_2(t)$, $\hat{\tau}_3(t)$ in Lemmas \ref{lem:1}-\ref{lem:4}.

\begin{lemma} \label{lem:1}
Under Conditions \ref{ass:rand_cens}, \ref{cond:1a}, we have that $\hat{\tau}_0(t)$ is uniformly consistent to $\tau_{\text{sp}}(t)$ for $t\in [0,t_0]$.
\end{lemma}

\begin{proof}[Proof of Lemma \ref{lem:1}] By definition,
\begin{eqnarray*}
\hat{\tau}_0(t)&=&n_1^{-1}\sum_{i:Z_i=1}\frac{R_i(t) \mathbbm{1}(T_i>t)}{\hat{S}_C(t|Z_i,X_i)}-n_0^{-1}\sum_{i:Z_i=0}\frac{R_i(t) \mathbbm{1}(T_i>t)}{\hat{S}_C(t|Z_i,X_i)}\\
&=&n_1^{-1}\sum_{i:Z_i=1}\frac{R_i(t) \mathbbm{1}(T_i>t)}{S_C(t|Z_i,X_i)}-n_0^{-1}\sum_{i:Z_i=0}\frac{R_i(t) \mathbbm{1}(T_i>t)}{S_C(t|Z_i,X_i)}+R_n(t),
\end{eqnarray*}

where the residual term \[R_n(t)=R_{n1}(t)-R_{n0}(t), ~\textnormal{with}~ R_{nz}(t)=n_z^{-1}\sum_{i:Z_i=z}\frac{(\pi_i(t)-\hat{\pi}_i(t))R_i(t)\mathbbm{1}(T_i> t)}{S_C(t|Z_i,X_i)\hat{S}_C(t|Z_i,X_i)}.\] 

Under Condition \ref{ass:rand_cens}, $S_C(t|Z_i,X_i)\geq \delta>0$, 
\begin{eqnarray*}
\sup_{t\in [0,t_0]}|R_{n}(t)|&\leq& \sup_{i}\frac{|S_C(t|Z_i,X_i)-\hat{S}_C(t|Z_i,X_i)|R_i(t)\mathbbm{1}(T_i> t)}{S_C(t|Z_i,X_i)\hat{S}_C(t|Z_i,X_i)}\\
&\leq & \sup_{\substack{z\in \{0,1\},\\ x\in \mathbb{X},\\t\in [0,t_0]}} |S_C(t|z,x)-\hat{S}_C(t|z,x)| \frac{2}{\delta^2}
\end{eqnarray*}
under the event 
\begin{equation}
   A_n=\left\{\sup_{t\in [0,t_0]}\sup_i|\hat{S}_C(t|Z_i,X_i)-S_C(t|Z_i,X_i)|\leq \frac{\delta}{2} \right\}. 
\end{equation}
 Given Condition \ref{cond:1a}, we have 
\begin{eqnarray*}
    \mathbb{P}(A_n^c)&\leq &\mathbb{P}\left(\sup_{\substack{z\in \{0,1\},\\ x\in \mathbb{X},\\t\in [0,t_0]}}|\hat{S}_C(t|z,x)-S_C(t|z,x)|> \frac{\delta}{2}\right)\rightarrow 0
\end{eqnarray*}
So we conclude $|R_n(t)|=o_p(1)$ uniformly. So using uniform law of large number (ULLN), we have
\begin{eqnarray*}
\hat{\tau}_0(t)&=& \mathbb{E}\left\{\frac{R_i(t) \mathbbm{1}(T_i>t)}{S_C(t|Z_i,X_i)}|Z_i=1\right\}- \mathbb{E}\left\{\frac{R_i(t) \mathbbm{1}(T_i>t)}{S_C(t|Z_i,X_i)}|Z_i=0\right\}+o_p(1)\\
&=&\tau_{\text{sp}}(t)+o_p(1),
\end{eqnarray*}
where the last equality is because of Condition \ref{ass:rand_cens}. This conclude the proof of consistency of $\hat{\tau}_0(t)$.
\end{proof}

\begin{lemma} \label{lem:2} Under Condition \ref{cond:2}, we have $\hat{\tau}_1(t)$ is uniformly consistency to $\tau_{\text{sp}}(t)$ for all $t\in [0,t_0]$.
\end{lemma}

\begin{proof}[Proof of Lemma \ref{lem:2}] We have
\begin{eqnarray*}
\hat{\tau}_1(t)&=&n^{-1}\sum_{i=1}^n\left \{\hat{\mu}^{(1,-i)}(t,X_i)-\hat{\mu}^{(0,-i)}(t,X_i)\right\}\\
&=&n^{-1}\sum_{i=1}^n \{\mu^{(1)}(t,X_i)-\mu^{(0)}(t,X_i)\}+R^{(1)}_{n}(t)\\
&=&\tau_{\text{sp}}(t)+R^{(1)}_{n}(t)
\end{eqnarray*}
where $R^{(1)}_{n}(t)=R^{(1)}_{n1}(t)-R^{(1)}_{n0}(t)$ with $R^{(1)}_{nz}(t)=n^{-1}\sum_{i=1}^n\left \{\mu^{(z)}(t,X_i)-\hat{\mu}^{(z,-i)}(t,X_i)\right\}$ for $z=0,1$. By Condition \ref{cond:2}, we have
\begin{align*}
\sup_{t\in [0,t_0]}|R^{(1)}_{nz}(t)|\leq \sup_{t\in [0,t_0]}\sup_i |\mu^{(z)}(t,X_i)-\hat{\mu}^{(z,-i)}(t,X_i)| \\ \leq \sup_{\substack{z\in \{0,1\},\\ x\in \mathbb{X},\\t\in [0,t_0]}} |\mu^{(z)}(t,x)-\hat{\mu}^{(z,-i)}(t,x)|\stackrel{p}{\rightarrow} 0.
\end{align*}
So we have $\hat{\tau}_1(t)=\tau_{\text{sp}}(t)+o_p(1)$ which conclude the proof of consistency of $\hat{\tau}_1(t)$.
\end{proof}

\begin{lemma} \label{lem:3} Under Conditions \ref{ass:rand_cens}, \ref{cond:1a}, and \ref{cond:2}, we have $\hat{\tau}_2(t)$ is uniformly consistent to $\tau_{\text{sp}}(t)$ for $t\in [0,t_0]$.
\end{lemma}

\begin{proof}[Proof of Lemma \ref{lem:3}] We have
\begin{eqnarray*}
\hat{\tau}_2(t)&=&\hat{\tau}_0(t)+\hat{\tau}_1(t)+R^{(2)}(t)\\
&=&\tau_{\text{sp}}(t)+o_p(1)+\tau_{\text{sp}}(t)+o_p(1)+R^{(2)}(t)
\end{eqnarray*}
where by Condition \ref{cond:2}
\begin{eqnarray*}
R^{(2)}(t)&=&-n_1^{-1}\sum_{i:Z_i=1}\hat{\mu}^{(1,-i)}(t,X_i)+n_0^{-1}\sum_{i:Z_i=0}\hat{\mu}^{(0,-i)}(t,X_i)\\
&=&-n_1^{-1}\sum_{i:Z_i=1}\mu^{(1)}(t,X_i)+n_0^{-1}\sum_{i:Z_i=0}\mu^{(0)}(t,X_i)+o_p(1)\\
&=& \mathbb{E}[-n_1^{-1}\sum_{i:Z_i=1}\mu^{(1)}(t,X_i)]+\mathbb{E}[n_0^{-1}\sum_{i:Z_i=0}\mu^{(0)}(t,X_i)]+o_p(1)\\
&=& -\tau_{\text{sp}}(t)+o_p(1).
\end{eqnarray*}
Therefore,
\begin{eqnarray*}
\hat{\tau}_2(t)&=&2\tau_{\text{sp}}(t)+o_p(1)-\tau_{\text{sp}}(t)+o_p(1)=\tau_{\text{sp}}(t)+o_p(1),
\end{eqnarray*}
which conclude the proof of consistency of $\hat{\tau}_2(t)$.
\end{proof}

\begin{lemma} \label{lem:4} Under Conditions \ref{ass:rand_cens}, \ref{cond:1a}, \ref{cond:2}, we have $\hat{\tau}_3(t)$ is uniformly consistent to $\tau_{\text{sp}}(t)$ for $t\in [0,t_0]$.\end{lemma}

\begin{proof}[Proof of Lemma \ref{lem:4}] Notice that $|\mathbbm{1}(T_i>t)-\hat{\mu}^{(1,-i)}(t,X_i)|\leq 1$, so similarly as proof of Lemma \ref{lem:1}, we have under Condition \ref{ass:rand_cens} and \ref{cond:1a},
\begin{eqnarray*}
&&\sup_{t\in [0,t_0]}|n_z^{-1}\sum_{i:Z_i=z}\frac{(\pi_i(t)-\hat{\pi}_i(t))R_i(t)\left\{\mathbbm{1}(T_i>t)-\hat{\mu}^{(1,-i)}(t,X_i)\right\}}{\pi_i(t)\hat{\pi}_i(t)}|\\
&\leq& \sup_{z,x}\sup_{t\in [0,t_0]} |S_C(t|Z_i,X_i)-\hat{S}_C(t|Z_i,X_i)| \frac{2}{\delta^2}\rightarrow 0\\
\end{eqnarray*}
under event $A_n$. So we have
\begin{eqnarray*}
\hat{\tau}_3(t)&=&\hat{\tau}_1(t)+n_1^{-1}\sum_{i:Z_i=1} \frac{R_i(t)\left\{\mathbbm{1}(T_i>t)-\hat{\mu}^{(1,-i)}(t,X_i)\right\}}{S_C(t\wedge T_i|Z_i,X_i)}\\
&&-n_0^{-1}\sum_{i:Z_i=0} \frac{R_i(t)\left\{\mathbbm{1}(T_i>t)-\hat{\mu}^{(0,-i)}(t,X_i)\right\}}{S_C(t\wedge T_i|Z_i,X_i)}+o_p(1)\\
&=&\hat{\tau}_1(t)+n_1^{-1}\sum_{i:Z_i=1} \frac{R_i(t)\left\{\mathbbm{1}(T_i>t)-\mu^{(1)}(t,X_i)\right\}}{S_C(t\wedge T_i|Z_i,X_i)}\\
&&-n_0^{-1}\sum_{i:Z_i=0} \frac{R_i(t)\left\{\mathbbm{1}(T_i>t)-\mu^{(0)}(t,X_i)\right\}}{S_C(t\wedge T_i|Z_i,X_i)}+o_p(1)\\
&=&\tau_{\text{sp}}(t)+ \mathbb{E} \left[\frac{R_i(t)\left\{\mathbbm{1}(T_i>t)-\hat{\mu}^{(1,-i)}(t,X_i)\right\}}{S_C(t\wedge T_i|Z_i,X_i)}|Z_i=1\right]\\ &&-\mathbb{E}\left[\frac{R_i(t)\left\{\mathbbm{1}(T_i>t)-\hat{\mu}^{(0,-i)}(t,X_i)\right\}}{S_C(t\wedge T_i|Z_i,X_i)}|Z_i=0\right]+o_p(1)\\
&=& \tau_{\text{sp}}(t)+o_p(1)
\end{eqnarray*}
This finish the proof of consistency of $\hat{\tau}_3(t)$.
\end{proof}


\subsection*{Proof of Theorem 2:}
From the definition of $\hat{\tau}_0(t)$,
\begin{eqnarray*}
&&\sqrt{n}(\hat{\tau}_0(t)-\tau_{\text{sp}}(t))\\
&=&\sqrt{n}\left\{n_1^{-1}\sum_{i:Z_i=1}\frac{R_i(t) \mathbbm{1}(T_i>t)}{S_C(t|Z_i,X_i)}-n_0^{-1}\sum_{i:Z_i=0}\frac{R_i(t) \mathbbm{1}(T_i>t)}{S_C(t|Z_i,X_i)}-\tau_{\text{sp}}(t)\right\}\\
&&+\sqrt{n}\cdot n_1^{-1}\sum_{i:Z_i=1}\frac{\hat{S}_C(t|Z_i,X_i)-S_C(t|Z_i,X_i)}{S_C(t|Z_i,X_i)^2}R_i(t) \mathbbm{1}(T_i>t)\\
&&-\sqrt{n}\cdot n_0^{-1}\sum_{i:Z_i=0}\frac{\hat{S}_C(t|Z_i,X_i)-S_C(t|Z_i,X_i)}{S_C(t|Z_i,X_i)^2}R_i(t) \mathbbm{1}(T_i>t)+R^{(4)}_n(t),
\end{eqnarray*}
where $R^{(4)}_n(t)=R^{(4)}_{n1}(t)-R^{(4)}_{n0}(t)$ with $R^{(4)}_{nz}(t)=\sqrt{n}\cdot n_z^{-1}\sum_{i:Z_i=z}\frac{[\hat{S}_C(t|Z_i,X_i)-S_C(t|Z_i,X_i)]^2}{\hat{S}_C(t|Z_i,X_i)S_C(t|Z_i,X_i)^2}$. So we have for $z \in \{0,1\}$,
\begin{eqnarray*}
\sup_{t\in [0,t_0]} |R^{(4)}_{nz}(t)|\leq \sup_{\substack{t\in [0,t_0]\\ x\in \mathbb{X}\\ z \in \{0,1\}}} \sqrt{n}\left[\hat{S}_C(t|x,z)-S_C(t|x,z)\right]^2\frac{2}{\delta^3} \stackrel{p}{\rightarrow} 0
\end{eqnarray*}
under $A_n$. So we have
{\small
\begin{eqnarray*}
&&\sqrt{n}(\hat{\tau}_0(t)-\tau_{\text{sp}}(t))\\
&=&\sqrt{n}\left\{n_1^{-1}\sum_{i:Z_i=1}\frac{R_i(t) \mathbbm{1}(T_i>t)}{S_C(t|Z_i,X_i)}-n_0^{-1}\sum_{i:Z_i=0}\frac{R_i(t) \mathbbm{1}(T_i>t)}{S_C(t|Z_i,X_i)}-\tau_{\text{sp}}(t)\right\}\\
&&+\sqrt{n}\cdot n_1^{-1}\sum_{i:Z_i=1}\frac{\hat{S}_C(t|Z_i,X_i)-S_C(t|Z_i,X_i)}{S_C(t|Z_i,X_i)^2}R_i(t) \mathbbm{1}(T_i>t)\\&&-\sqrt{n}\cdot n_0^{-1}\sum_{i:Z_i=0}\frac{\hat{S}_C(t|Z_i,X_i)-S_C(t|Z_i,X_i)}{S_C(t|Z_i,X_i)^2}R_i(t) \mathbbm{1}(T_i>t)+o_p(1)\\
&=&n^{-1/2}\sum_{i=1}^n \left[\left\{\frac{n_1}{n}Z_i-\frac{n_0}{n}(1-Z_i)\right\}^{-1}S_C^{-1}(t|Z_i,X_i)R_i(t)\mathbbm{1}(T_i>t)-\tau_{\text{sp}}(t)\right]\\
&&+n^{-1/2}\sum_{i=1}^n \left[\left\{\frac{n_1}{n}Z_i-\frac{n_0}{n}(1-Z_i)\right\}^{-1} S_C^{-2}(t|Z_i,X_i)R_i(t)\mathbbm{1}(T_i>t)(\hat{S}_C(t|Z_i,X_i)-S_C(t|Z_i,X_i))\right]+o_p(1)
\end{eqnarray*}}
Define 
\begin{eqnarray*}
U_{1i}(t)&=&\left\{\alpha Z_i-(1-\alpha)(1-Z_i)\right\}^{-1}S_C^{-1}(t|Z_i,X_i)R_i(t)\mathbbm{1}(T_i>t)-\tau_{\text{sp}}(t)
\end{eqnarray*}
then we have
{\small
\begin{eqnarray*}
&&\sqrt{n}(\hat{\tau}_0(t)-\tau_{\text{sp}}(t))\\
&=&n^{-1/2}\sum_{i=1}^n \left[U_{1i}(t)+\left\{\alpha Z_i-(1-\alpha)(1-Z_i)\right\}^{-1} S_C^{-2}(t|Z_i,X_i)R_i(t)\mathbbm{1}(T_i>t)(\hat{S}_C(t|Z_i,X_i)-S_C(t|Z_i,X_i))\right]+o_p(1)\\
&=&n^{-1/2}\sum_{i=1}^n \left[U_{1i}(t)+\left\{\alpha Z_i-(1-\alpha)(1-Z_i)\right\}^{-1} S_C^{-2}(t|Z_i,X_i)R_i(t)\mathbbm{1}(T_i>t)n^{-1}\sum_{j=1}^n\psi(t,Z_i,X_i;\mathcal{D}_j)\right]+o_p(1)\\
&=&n^{-1/2}\sum_{i=1}^n \left[U_{1i}(t)+n^{-1}\sum_{j=1}^n\left\{\alpha Z_j-(1-\alpha)(1-Z_j)\right\}^{-1} S_C^{-2}(t|X_j,Z_j)R_j(t)\mathbbm{1}(T_j>t)\psi(t,X_j,Z_j;\mathcal{D}_i)\right]+o_p(1)\\
\end{eqnarray*}}
Under Condition \ref{cond:1c}, we have

\begin{eqnarray*}
&&\sqrt{n}(\hat{\tau}_0(t)-\tau_{\text{sp}}(t))\\
&=&n^{-1/2}\sum_{i=1}^n \Big[U_{1i}(t)+\sum_{k=1}^Kn^{-1}\sum_{j=1}^n\left\{\alpha Z_j-(1-\alpha)(1-Z_j)\right\}^{-1}\cdot \\ && S_C^{-2}(t|X_j,Z_j)R_j(t)\mathbbm{1}(T_j>t)f_k(X_j,Z_j)\psi_{k}(t;\mathcal{D}_i)\Big]+o_p(1)\\
&=&n^{-1/2}\sum_{i=1}^n \Big[U_{1i}(t)+\sum_{k=1}^K\beta_{k}\psi_{k}(t;\mathcal{D}_i)\Big]+o_p(1)\\
&=&n^{-1/2}\sum_{i=1}^n \left[U_{1i}(t)+U_{2i}(t)\right]+o_p(1)
\end{eqnarray*}
where $U_{2i}(t)=\sum_{k=1}^K \beta_k \psi_k(t;\mathcal{D}_i)$ and $$\beta_k=\mathbb{E}\left[\left\{\alpha Z_j-(1-\alpha)(1-Z_j)\right\}^{-1} S_C^{-2}(t|X_j,Z_j)R_j(t)\mathbbm{1}(T_j>t)f_k(X_j,Z_j)\right].$$

Therefore, $\sqrt{n}(\hat{\tau}_0(t)-\tau_{\text{sp}}(t))\rightarrow U_0(t)$, where $U_0(t)$ is a mean 0 Gaussian process with covariance process
\begin{eqnarray*}
\Sigma_0(t,s)=Cov(U_0(t),U_0(s))=\lim_{n\rightarrow \infty} Cov(U_{1i}(t)+U_{2i}(t),U_{1i}(s)+U_{2i}(s)).
\end{eqnarray*}

Below we give an example assuming random censoring. In this case we can simply estimate $\hat{S}_C(t|X_j,Z_j)=\hat{S}_C(t)$ using Kaplan-Meier estimator. Condition \ref{cond:1c} is satisfied and specifically, we have 
\begin{eqnarray*}
\psi_i(t,x,z)=-S_C(t)\int_0^t \frac{1}{\mathbb{P}(T\wedge C\geq s)}dM_{Ci}(s)
\end{eqnarray*}
where $dM_{Ci}(t)=dN_{Ci}(t)-\lambda_c(s)\mathbbm{1}(T_i\wedge C_i\geq t)ds$ with $N_{Ci}(t)=\mathbbm{1}(T_i\wedge C_i\leq t,\Delta_i=0)$ and $\lambda_c(s)$ is the baseline hazard for the censoring process. Based on this, we have
\begin{eqnarray*}
&&\sqrt{n}(\hat{\tau}_0(t)-\tau_{\text{sp}}(t))=n^{-1/2}\sum_{i=1}^n \left[U_{1i}(t)+U_{2i}(t)\right]+o_p(1)
\end{eqnarray*}
where
\begin{eqnarray*}
U_{2i}(t)&=&-\mathbb{E}\left[\left\{\alpha Z_j-(1-\alpha)(1-Z_j)\right\}^{-1} S_C^{-1}(t)R_j(t)\mathbbm{1}(T_j>t)\right]\int_0^{t} \frac{1}{\mathbb{P}(T\wedge C\geq s)}dM_{Ci}(s)\\
&=&-\tau_{\text{sp}}(t)\int_0^{t} \frac{1}{\mathbb{P}(T\wedge C\geq s)}dM_{Ci}(s)
\end{eqnarray*}
Under the regularity conditions of the functional central limit theorem of \citet{pollard1982}, we have
\begin{eqnarray*}
\sqrt{n}\left\{\hat{\tau}_0(t)-\tau_{\text{sp}}(t)\right\}=n^{-1/2}\sum_{i=1}^n \left\{U_{1i}(t)+U_{2i}(t)\right\}+o_p(1)\rightarrow U_0(t)
\end{eqnarray*}
where $U_0(t)$ is a mean 0 Gaussian process with covariance process
\begin{eqnarray*}
\Sigma_0(t,s)=Cov(U_0(t),U_0(s))=\lim_{n\rightarrow \infty} Cov(U_{1i}(t)+U_{2i}(t),U_{1i}(s)+U_{2i}(s))
\end{eqnarray*}
which can be estimated by the empirical covariance
\begin{eqnarray*}
n^{-1}\sum_{i=1}^n(\hat{U}_{1i}(t)+\hat{U}_{2i}(t))(\hat{U}_{1i}(s)+\hat{U}_{2i}(s))
\end{eqnarray*}
where $\hat{U}_{1i}(t)$ and $\hat{U}_{2i}(t)$ are plug-in estimators for $U_{1i}(t)$, $U_{2i}(t)$, i.e.,
\begin{eqnarray*}
\hat{U}_{1i}(t)&=&\left\{\hat{\alpha}Z_i-(1-\hat{\alpha})(1-Z_i)\right\}^{-1}\hat{S}_C^{-1}(t)R_i(t)\mathbbm{1}(T_i>t)-\hat{\tau}_0(t)\\
\hat{U}_{2i}(t)&=&-\hat{\tau}_0(t)\int_0^t \frac{1}{n^{-1}\sum_{j=1}^n \mathbbm{1}(Y_j\geq s)}d\hat{M}_{Ci}(s)
\end{eqnarray*}
where $\hat{\alpha}=\frac{n_1}{n}$ and $\hat{M}_{Ci}(s)=\mathbbm{1}(Y_i\leq s, \Delta_i=0)-\int_0^s \mathbbm{1}(Y_i\geq u)d\hat{\Lambda}_C(u)$ where $\hat{\Lambda}_C(u)=-log(\hat{S}_C(u))$. 

Although the above derivation is for Kaplan-Meier estimator, as a remark, the asymptotic normality holds for other estimators as long as the influnce function can be written as $\psi(t,x,z;\mathcal{D}_i)=\sum_{k=1}^K f_k(x,z)\psi_k(t;\mathcal{D}_i)$, which is the case for stratified Kaplan Meier estimator and Cox regression based estimators. Under such formula, the assymptotic normality of $\sqrt{n}(\hat{\tau}_0(t)-\tau_{\text{sp}}(t))$ still holds, but the influence function and covariance process of the Gaussian process may be complex. In practice, we can use bootstrap to estimate the covariance.

\subsection*{Proof of Theorem 3:}
We have an expansion of $\hat{\tau}_2(t)$ as follow:
\begin{eqnarray*}
\hat{\tau}_2(t)&=&n^{-1}\sum_{i=1}^n(\mu^{(1)}(t,X_i)-\mu^{(0)}(t,X_i))\\
&&+n_1^{-1}\sum_{i:Z_i=1} \left\{\frac{R_i(t)\mathbbm{1}(T_i>t)}{\hat{\pi}_i(t)}-\mu^{(1)}(t,X_i)\right\}-n_0^{-1}\sum_{i:Z_i=0}\left\{\frac{R_i(t)\mathbbm{1}(T_i>t)}{\hat{\pi}_i(t)}-\mu^{(0)}(t,X_i)\right\}+r_2(t)
\end{eqnarray*}
where the residual $r_2(t)$ has the expression
\begin{eqnarray*}
r_2(t)=\sum_{i=1}^n \frac{(-1)^{Z_i}}{n_{Z_i}}\left[\frac{n_0}{n}\left\{\hat{\mu}^{(1,-i)}(t,X_i)-\mu^{(1)}(t,X_i)\right\}+\frac{n_1}{n}\left\{\hat{\mu}^{(0,-i)}(t,X_i)-\mu^{(0)}(t,X_i)\right\}\right].
\end{eqnarray*}

Here we will show the residual term $r_2(t)$ is asymptotically negligible under Jackknife compatible condition. We first define a ``leave-two-out'' approximation of $r_2(t)$,
\begin{eqnarray*}
r_{22}(t)=\sum_{i=1}^n \frac{(-1)^{Z_i}}{n_{0}n_{1}}\sum_{j:Z_i\neq Z_j}\left[\frac{n_0}{n}\left\{\hat{\mu}^{(1,-\{i,j\})}(t,X_i)-\mu^{(1)}(t,X_i)\right\}+\frac{n_1}{n}\left\{\hat{\mu}^{(0,-\{i,j\})}(t,X_i)-\mu^{(0)}(t,X_i)\right\}\right]
\end{eqnarray*}
where $\hat{\mu}^{(z,-\{i,j\})}$ are predictions obtained without either the $i$th or the $j$th individuals for training model (i.e., remove one observation from both treatment and control group). The randomization guarantee that $\hat{\mu}^{(z,-\{i,j\})}$ is independent of $Z_i$ conditionally on $n_1$, So we have $\mathbb{E}[r_{22}(t)]=0$ and 
{\small
\begin{eqnarray*}
\mathbb{E}[r_{22}^2(t)]&=&n\mathbb{E}\left\{\frac{\left[\sum_{j:Z_i\neq Z_j}\left[\frac{n_0}{n}\left\{\hat{\mu}^{(1,-\{i,j\})}(t,X_i)-\mu^{(1)}(t,X_i)\right\}+\frac{n_1}{n}\left\{\hat{\mu}^{(0,-\{i,j\})}(t,X_i)-\mu^{(0)}(t,X_i)\right\}\right]\right]^2}{n_0^2n_1^2}\right\}\\
&=&\mathbb{E}\left\{\frac{\left[\sum_{j:Z_i\neq Z_j}\left[n_0\left\{\hat{\mu}^{(1,-\{i,j\})}(t,X_i)-\mu^{(1)}(t,X_i)\right\}+n_1\left\{\hat{\mu}^{(0,-\{i,j\})}(t,X_i)-\mu^{(0)}(t,X_i)\right\}\right]\right]^2}{n_0^2n_1^2}\right\}\\
&\leq & 2\mathbb{E}\left[\frac{\mathbb{E}\left[\left\{\sum_j(\hat{\mu}^{(1,-\{i,j\})}(t,X_i)-\mu^{(1)}(t,X_i))\right\}^2|n_1\right]}{n_1^2}+\frac{\mathbb{E}\left[\left\{\sum_j(\hat{\mu}^{(0,-\{i,j\})}(t,X_i)-\mu^{(0)}(t,X_i))\right\}^2|n_0\right]}{n_0^2}\right]\\
&\leq & 2n\mathbb{E}\left[\frac{\mathbb{E}\left[\sum_j\left\{\hat{\mu}^{(1,-\{i,j\})}(t,X_i)-\mu^{(1)}(t,X_i)\right\}^2|n_1\right]}{n_1^2}+\frac{\mathbb{E}\left[\sum_j\left\{\hat{\mu}^{(0,-\{i,j\})}(t,X_i)-\mu^{(0)}(t,X_i)\right\}^2|n_0\right]}{n_0^2}\right]\\
&\leq&2n\mathbb{E}\left\{\frac{a(n_1)}{n_1^2}+\frac{a(n_0)}{n_0^2}\right\}, \quad \text{--- because of Condition \ref{cond:5},}\\
&\leq &2n^{-1}\max_{z\in \{0,1\}} a(n_z)(\min\{\alpha,1-\alpha\})^{-2}\\
&=&o(n^{-1}), \quad \text{--- since under Condition \ref{cond:5} we have $a(n_z)=o(1)$.}
\end{eqnarray*} }
Therefore, $r_{22}(t)=o_p(n^{-1/2})$. Also, from Condition \ref{cond:4}, we have
\begin{eqnarray*}
\frac{1}{2}\mathbb{E}\left[\left\{r_2(t)-r_{22}(t)\right\}^2|n_1\right]\leq \frac{n_0^2a(n_1)}{n^2n_1}+\frac{n_1^2a(n_0)}{n^2n_0}=o_p(1/n).
\end{eqnarray*}
So we get $r_{2}(t)-r_{22}(t)=o_p(n^{-1/2})$ which combine with $r_{22}(t)=o_p(n^{-1/2})$ give us $r_2(t)=o_p(n^{-1/2})$ is negligible.

Now using the results for the asymptotic normality of $\hat{\tau}_0(t)$, we can directly write out that under random censoring
\begin{eqnarray*}
\sqrt{n}(\hat{\tau}_2(t)-\tau_{\text{sp}}(t))=n^{-1/2}\sum_i (U_{1i}(t)+U_{2i}(t)+U_{3i}(t))
\end{eqnarray*}
where
\begin{eqnarray*}
U_{3i}(t)=\mu^{(1)}(t,X_i)-\mu^{(0)}(t,X_i)-\left\{\alpha Z_i-(1-\alpha)(1-Z_i)\right\}^{-1}\mu^{(Z_i)}(t, X_i)
\end{eqnarray*}

So under the regularity conditions of the functional central limit theorem of \citet{pollard1982}, we have
\begin{eqnarray*}
\sqrt{n}\left\{\hat{\tau}_2(t)-\tau_{\text{sp}}(t)\right\}=n^{-1/2}\sum_{i=1}^n \left\{U_{1i}(t)+U_{2i}(t)+U_{3i}(t)\right\}+o_p(1)\Rightarrow U_2(t)
\end{eqnarray*}
where $U_2(t)$ is a mean 0 Gaussian process with covariance process
\begin{eqnarray*}
\Sigma_2(t,s)=Cov(U_2(t),U_2(s))=\lim_{n\rightarrow \infty} Cov(U_{1i}(t)+U_{2i}(t)+U_{3i}(t),U_{1i}(s)+U_{2i}(s)+U_{3i}(s))
\end{eqnarray*}
which can be estimated by the empirical covariance
\begin{eqnarray*}
n^{-1}\sum_{i=1}^n(\hat{U}_{1i}(t)+\hat{U}_{2i}(t)+\hat{U}_{3i}(t))(\hat{U}_{1i}(s)+\hat{U}_{2i}(s)+\hat{U}_{3i}(s))
\end{eqnarray*}
where $\hat{U}_{3i}$ are plug-in estimator for $U_{3i}$, i.e.,
\begin{eqnarray*}
\hat{U}_{3i}(t)&=&\hat{\mu}^{(1,-i)}(t,X_i)-\hat{\mu}^{(0,-i)}(t,X_i)-\left\{\hat{\alpha} Z_i-(1-\hat{\alpha})(1-Z_i)\right\}^{-1}\hat{\mu}^{(Z_i,-i)}(t, X_i).
\end{eqnarray*}

We rewrite $\hat{\tau}_3(t)$ as
\begin{eqnarray*}
\hat{\tau}_3(t)&=&n^{-1}\sum_{i=1}^n\left\{\mu^{(1)}(t,X_i)-\mu^{(0)}(t,X_i)\right\}+n_1^{-1}\sum_{i:Z_i=1} \left[\frac{R_i(t)\left\{\mathbbm{1}(T_i>t)-\mu^{(1)}(t,X_i)\right\}}{\hat{\pi}_i(t)}\right]\\
&&-n_0^{-1}\sum_{i:Z_i=0}\left[\frac{R_i(t)\left\{\mathbbm{1}(T_i>t)-\mu^{(0)}(t,X_i)\right\}}{\hat{\pi}_i(t)}\right]+r_2(t)+r_3(t)
\end{eqnarray*}
where the additional residual $r_3(t)$ has the expression
\begin{eqnarray*}
r_3(t)&=&\sum_{i=1}^n (-1)^{Z_i}n_{Z_i}^{-1} \left\{1-\frac{R_i(t)}{\hat{\pi}_i(t)}\right\}\left\{\hat{\mu}^{(Z_i,-i)}(t,X_i)-\mu^{(Z_i)}(t,X_i)\right\}\\
&=&\sum_{i=1}^n (-1)^{Z_i}n_{Z_i}^{-1} \left\{1-\frac{R_i(t)}{\pi_i(t)}\right\}\left\{\hat{\mu}^{(Z_i,-i)}(t,X_i)-\mu^{(Z_i)}(t,X_i)\right\}+o_p(n^{-1/2})\\
\end{eqnarray*}
Using the same argument as $r_2(t)$, we have $r_3(t)$ is negligible. So we have 
\begin{eqnarray*}
\sqrt{n}\left\{\hat{\tau}_3(t)-\tau_{\text{sp}}(t)\right\}=n^{-1/2}\sum_{i=1}^n \left\{U_{1i}(t)+U_{2i}(t)+U_{4i}(t)\right\}+o_p(1)\Rightarrow U_3(t)
\end{eqnarray*}
where
\begin{eqnarray*}
U_{4i}(t)=\mu^{(1)}(t,X_i)-\mu^{(0)}(t,X_i)-\left\{\alpha Z_i-(1-\alpha)(1-Z_i)\right\}^{-1}\frac{R_i(t)\mu^{(Z_i)}(t, X_i)}{\pi_i(t)}
\end{eqnarray*}

So under regularity, using functional central limit theorem of \citet{pollard1982}, we have
\begin{eqnarray*}
\sqrt{n}\left\{\hat{\tau}_3(t)-\tau_{\text{sp}}(t)\right\}=n^{-1/2}\sum_{i=1}^n \left\{U_{1i}(t)+U_{2i}(t)+U_{4i}(t)\right\}+o_p(1)\Rightarrow U_3(t)
\end{eqnarray*}
where $U_3(t)$ is a mean 0 Gaussian process with covariance process
\begin{eqnarray*}
\Sigma_3(t,s)=Cov(U_3(t),U_3(s))=\lim_{n\rightarrow \infty} Cov(U_{1i}(t)+U_{2i}(t)+U_{4i}(t),U_{1i}(s)+U_{2i}(s)+U_{4i}(s))
\end{eqnarray*}
which can be estimated by the empirical covariance
\begin{eqnarray*}
n^{-1}\sum_{i=1}^n(\hat{U}_{1i}(t)+\hat{U}_{2i}(t)+\hat{U}_{4i}(t))(\hat{U}_{1i}(s)+\hat{U}_{2i}(s)+\hat{U}_{4i}(s))
\end{eqnarray*}


where $\hat{U}_{4i}$ are plut-in estimator for $U_{4i}$, i.e.,
\begin{eqnarray*}
\hat{U}_{4i}(t)&=&\hat{\mu}^{(1,-i)}(t,X_i)-\hat{\mu}^{(0,-i)}(t,X_i)-\left\{\hat{\alpha} Z_i-(1-\hat{\alpha})(1-Z_i)\right\}^{-1}\frac{R_i(t)\hat{\mu}^{(Z_i,-i)}(t, X_i)}{\hat{\pi}_i(t)}.
\end{eqnarray*}
\subsection*{Proof of Theorem 4:}

We revisit the asymptotic covariance process for $\hat{\tau}_0(t)$ and $\hat{\tau}_2(t)$.

For $\hat{\tau}_0(t)$, the covariance process $\Sigma_0(t,s)=\Sigma_{01}(t,s)+\Sigma_{02}(t,s)+\Sigma_{03}(t,s)+\Sigma_{03}(s,t)$ where
\begin{eqnarray*}
\Sigma_{01}(t,s)&=&\lim_{n\rightarrow \infty}Cov\left\{U_{1i}(t),U_{1i}(s)\right\}\\
&=&\alpha^{-1}\left\{\frac{\mu^{(1)}(t\vee s)}{S_C(t\wedge s)}-\mu^{(1)}(t)\mu^{(1)}(s)\right\}+(1-\alpha)^{-1}\left\{\frac{\mu^{(0)}(t\vee s)}{S_C(t\wedge s)}-\mu^{(0)}(t)\mu^{(0)}(s)\right\}
\end{eqnarray*}
\begin{eqnarray*}
\Sigma_{02}(t,s)&=&\lim_{n\rightarrow \infty}Cov(U_{2i}(t),U_{2i}(s))\\
&=&\tau_{\text{sp}}(t)\tau_{\text{sp}}(s)E\int_0^{t\wedge s}\frac{dN_C(u)}{\left\{\mathbb{P}(Y\geq u)\right\}^2}\\
&=&\tau_{\text{sp}}(t)\tau_{\text{sp}}(s)\int_0^{t\wedge s}\frac{\lambda_C(u)Pr(Y>u)du}{\left\{\mathbb{P}(Y\geq u)\right\}^2}\\
&=&\tau_{\text{sp}}(t)\tau_{\text{sp}}(s)\int_0^{t\wedge s}\frac{\lambda_C(u)du}{\mathbb{P}(Y\geq u)}
\end{eqnarray*}
And we have
\begin{eqnarray*}
\Sigma_{03}(t,s)&=&\lim_{n\rightarrow \infty}Cov\left\{U_{1i}(t),U_{3i}(s)\right\}\\
&=&-\lim_{n\rightarrow \infty}\mathbb{E}\left[\left\{\alpha Z_i-(1-\alpha)(1-Z_i)\right\}^{-1}S_C^{-1}(t)R_i(t)\mathbbm{1}(T_i>t)\tau_{\text{sp}}(s)\int_0^s \frac{1}{\mathbb{P}(Y\geq u)}dM_{Ci}(u)\right]\\
&=&-\lim_{n\rightarrow \infty}\mathbb{E}\left[S_C^{-1}(t)R_i(t)\mathbbm{1}(T_i>t)\tau_{\text{sp}}(s)\int_0^s \frac{1}{\mathbb{P}(Y\geq u)}dM_{Ci}(u)|Z_i=1\right]\\
&&+\lim_{n\rightarrow \infty}\mathbb{E}\left[S_C^{-1}(t)R_i(t)\mathbbm{1}(T_i>t)\tau_{\text{sp}}(s)\int_0^s \frac{1}{\mathbb{P}(Y\geq u)}dM_{Ci}(u)|Z_i=0\right]
\end{eqnarray*}
Notice we have
\begin{eqnarray*}
&&\lim_{n\rightarrow \infty}\mathbb{E}\left[S_C^{-1}(t)R_i(t)\mathbbm{1}(T_i>t)\tau_{\text{sp}}(s)\int_0^s \frac{1}{\mathbb{P}(Y\geq u)}dM_{Ci}(u)|Z_i=z\right]\\
&=&S_C^{-1}(t)\tau_{\text{sp}}(s)\lim_{n\rightarrow \infty}\mathbb{E}\left[\mathbbm{1}(Y_i>t)\int_0^s \frac{1}{\mathbb{P}(Y\geq u)}dM_{Ci}(u)|Z_i=z\right]\\
&=&-S_C^{-1}(t)\tau_{\text{sp}}(s)\lim_{n\rightarrow \infty}\mathbb{E}\left[\int_0^t(dN_i(u)+dN_i^C(u))\int_0^s \frac{1}{\mathbb{P}(Y\geq u)}dM_{Ci}(u)|Z_i=z\right]\\
&=&-S_C^{-1}(t)\tau_{\text{sp}}(s)\lim_{n\rightarrow \infty}\mathbb{E}\left[\int_0^{t\wedge s}\frac{1}{\mathbb{P}(Y\geq u)}dN_{Ci}(u)|Z_i=z\right]\\
&=&-S_C^{-1}(t)\tau_{\text{sp}}(s)\lim_{n\rightarrow \infty}\left[\int_0^{t\wedge s}\frac{1}{\mathbb{P}(Y\geq u)}\lambda_C(u)\mathbb{P}(Y>u|Z=z)du\right]
\end{eqnarray*}
So we have
\begin{eqnarray*}
\Sigma_{03}(t,s)&=&S_C^{-1}(t)\tau_{\text{sp}}(s)\lim_{n\rightarrow \infty}\left[\int_0^{t\wedge s}\frac{1}{\mathbb{P}(Y\geq u)}\lambda_C(u)(\mathbb{P}(Y>u|Z=1)-\mathbb{P}(Y>u|Z=0))\right]\\
&=&\tau_{\text{sp}}(t)\int_0^{t\wedge s}\frac{\lambda_C(u)\tau_{\text{sp}}(u)du}{\mathbb{P}(Y\geq u)}
\end{eqnarray*}


For $\hat{\tau}_2(t)$, we have $\Sigma_2(t,s)=\Sigma_{21}(t,s)+\Sigma_{02}(t,s)+\Sigma_{23}(t,s)+\Sigma_{23}(s,t)$
\begin{eqnarray*}
\Sigma_{21}(t,s)&=&\lim_{n\rightarrow \infty}Cov\left\{U_{1i}(t)+U_{3i}(t),U_{1i}(s)+U_{3i}(s)\right\}\\
&=&n^{-1}\sum_{i=1}^n\left[\alpha^{-1}\left\{\frac{\mu_i^{(1)}(t\vee s)}{S_C(t\wedge s)}-\mu_i^{(1)}(t)\mu_i^{(1)}(s)\right\}+(1-\alpha)^{-1}\left\{\frac{\mu_i^{(0)}(t\vee s)}{S_C(t\wedge s)}-\mu_i^{(0)}(t)\mu_i^{(0)}(s)\right\}\right]
\end{eqnarray*}
and 
\begin{eqnarray*}
\Sigma_{23}(t,s)&=&\lim_{n\rightarrow \infty}Cov\left\{U_{1i}(t)+U_{3i}(t),U_{2i}(s)\right\}\\
&=&\Sigma_{03}(t,s)+\lim_{n\rightarrow \infty}\mathbb{E}\left[\left\{\alpha Z_i-(1-\alpha)(1-Z_i)\right\}^{-1}\mu^{Z_i}(t,X_i)\tau_{\text{sp}}(s)\int_0^s \frac{1}{\mathbb{P}(Y\geq u)}dM_{Ci}(u)\right]\\
&=&\Sigma_{03}(t,s)
\end{eqnarray*}


Notice that $\mu^{(z)}(t)=n^{-1}\sum_{i=1}^n \mu^{(z)}(t, X_i)$, so we have 
\begin{eqnarray*}
&& \Sigma_{01}(t,s)-\Sigma_{21}(t,s)\\
&=&\lim_{n\rightarrow \infty} \alpha^{-1}\left\{n^{-1}\sum_{i=1}^n\mu^{(1)}(t, X_i)\mu^{(1)}(s, X_i)-\mu^{(1)}(t)\mu^{(1)}(s)\right\}\\
&&+(1-\alpha)^{-1}\left\{n^{-1}\sum_{i=1}^n\mu^{(0)}(t, X_i)\mu^{(0)}(s, X_i)-\mu^{(0)}(t)\mu^{(0)}(s)\right\}\\
&=&\alpha^{-1}Cov\left\{\mu^{(1)}(t, X),\mu^{(1)}(s, X)\right\}+(1-\alpha)^{-1}Cov\left\{\mu^{(0)}(t, X),\mu^{(0)}(s, X)\right\}
\end{eqnarray*}
which is sum of two semi-positive definite covariance process and thus is semi-positive definite. Therefore, $\hat{\tau}_2$ is always more asymptotically efficient than $\hat{\tau}_0$. 

To compare $\hat{\tau}_3$ and $\hat{\tau}_2$, we have that the $\hat{\tau}_3$ is the augmented IPCW estimator with the augmented term
\begin{eqnarray*}
U_{3i}(t)-U_{4i}(t)=\left\{\alpha Z_i-(1-\alpha)(1-Z_i)\right\}^{-1}\mu^{Z_i}(t,X_i)\frac{R_i(t)-\pi_i(t)}{\pi_i(t)}.
\end{eqnarray*}
The term above is independent of $U_{1i}(t)$ and is positively correlated with $U_{2i}(t)$ and $U_{4i}(t)$. So we have
\begin{eqnarray*}
&& Cov(U_3(t),U_3(s))\\
&=&Cov(U_{1i}(t)+U_{2i}(t)+U_{4i}(t),U_{1i}(s)+U_{2i}(s)+U_{4i}(s))\\
&\leq& Cov((U_{1i}(t)+U_{2i}(t)+U_{4i}(t))+(U_{3i}(t)-U_{4i}(t)),(U_{1i}(s)+U_{2i}(s)+U_{4i}(s))+(U_{3i}(s)-U_{4i}(s)))\\
&=&Cov(U_2(t),U_2(s)).
\end{eqnarray*}

Therefore, we have $\hat{\tau}_3$ is asymptotically more efficient than $\hat{\tau}_2$.

\newpage
\subsection*{Appendix B}
Additional simulation figures.
\begin{figure}[h]
\includegraphics[width=5in]{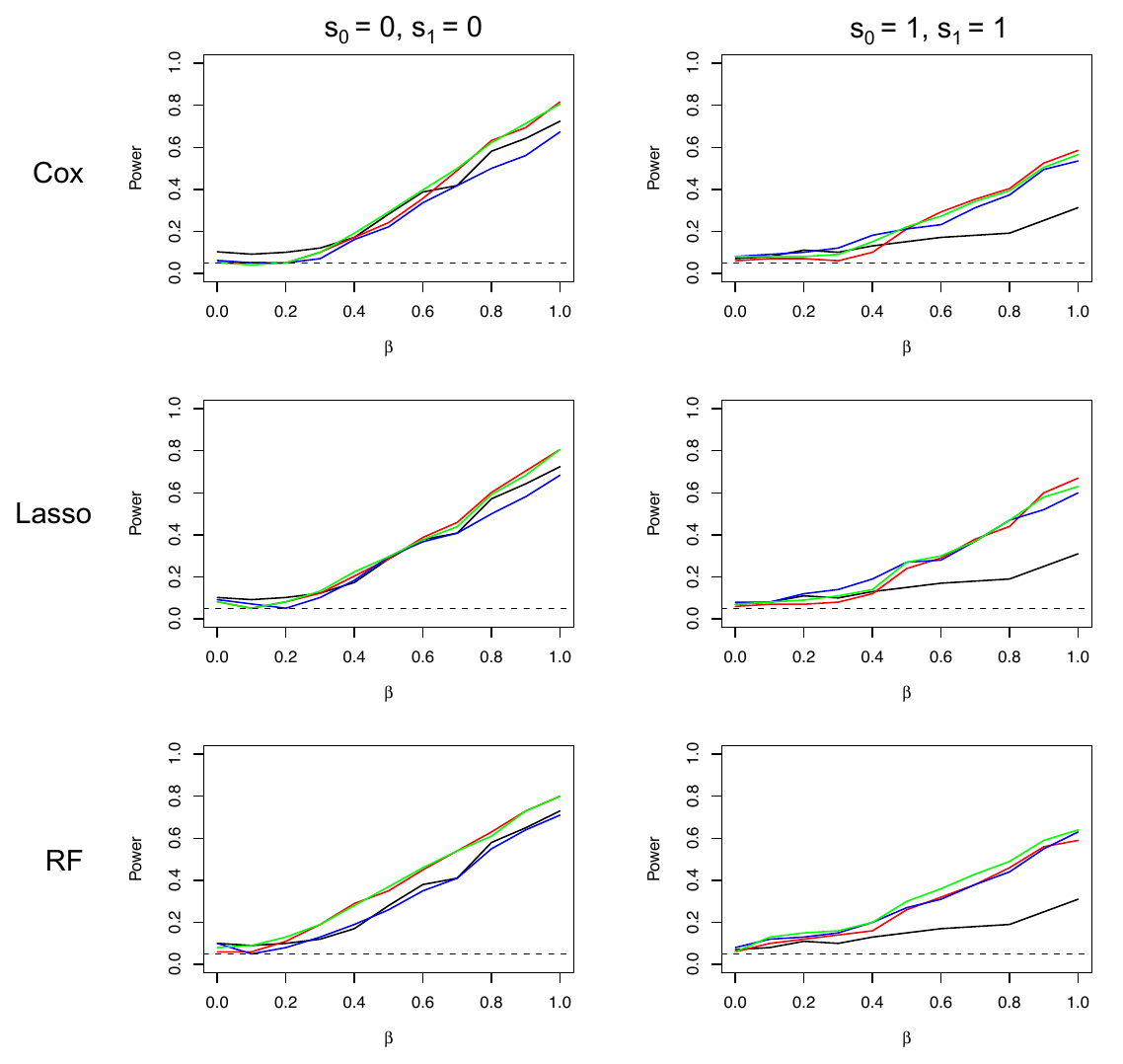}\\
\textbf{Figure S1:} Power curve for different estimators under low dimensional setting ($p=10$, $k=10$). Different estimators are presented with different colors as below: $\hat{\tau}_0$,black; $\hat{\tau}_1$,red; $\hat{\tau}_2$,blue; $\hat{\tau}_3$,green.\\
\label{figS1}
\end{figure}

\begin{figure}
\includegraphics[width=5in]{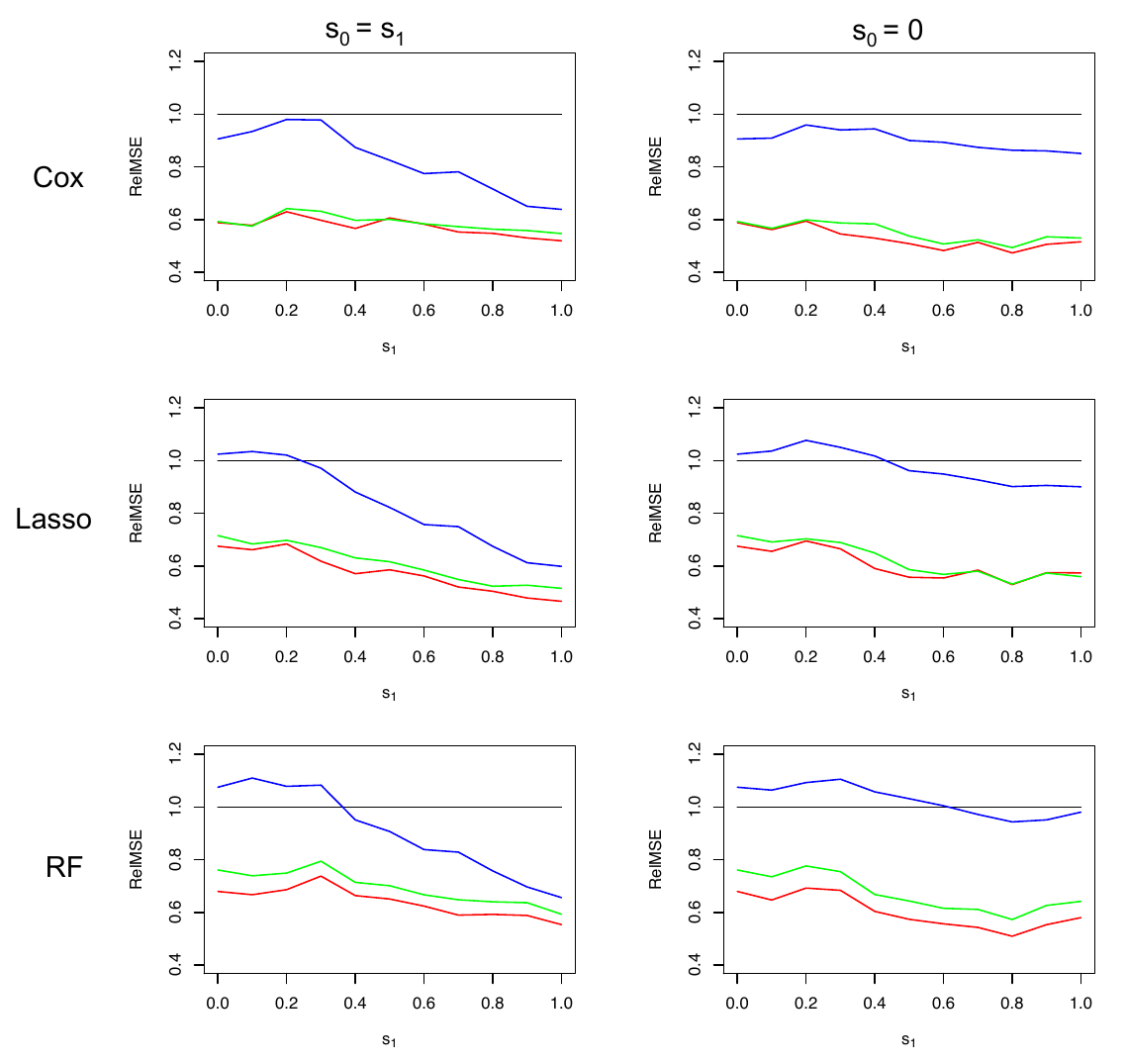}\\
\textbf{Figure S2:} Relative efficiency with the change of covariate effect for different estimators under low dimensional setting ($p=10$, $k=10$) when $\beta=0.5$. Different estimators are presented with different colors as below: $\hat{\tau}_0$,black; $\hat{\tau}_1$,red; $\hat{\tau}_2$,blue; $\hat{\tau}_3$,green.\\
\label{figS2}
\end{figure}

\newpage

\subsection*{Appendix C}
List of Cytogenetic abnormalities.
\begin{table}
{\small
\textbf{Table S1:} Cytogenetic abnormalities\\
\label{tabS1}
\begin{tabular}{|l|l|}\hline
Cyto Risk & Description\\\hline
\textit{del1q} & A portion of chromosome deleted from long arm (q) of chromosome 1\\\hline
\textit{del6q} & A portion of chromosome deleted from long arm (q) of chromosome 6\\\hline
\textit{del11q} & A portion of chromosome deleted from long arm (q) of chromosome 11\\\hline
 \textit{del17p} & A portion of chromosome deleted from short arm (p) of chromosome 17\\\hline
\textit{trisx}   & Extra copy of chromosome X\\\hline
 \textit{tris6} & Extra copy of chromosome 6\\\hline
  \textit{tris8} & Extra copy of chromosome 8\\\hline
 \textit{tris10} & Extra copy of chromosome 10\\\hline
 \textit{tris22} & Extra copy of chromosome 22\\\hline
 \textit{t411} & Translocation (4;11): \\
                             & portion of chromosome on 4 and 11 switched location and transferred to each other's\\\hline
 \textit{t119} & Translocation (1;19)\\\hline
 \textit{t1011} & Translocation (10;11)\\\hline
 \textit{t1119} & Translocation (11;19)\\\hline
 \textit{m13} & Monosomy 13: missing one copy of chromosome 13\\\hline
 \textit{mk} & Two or more autosomal monosomies or one autosomal monosomy associated with\\
                          &  at least one structural abnormality\\ 
                          & The most frequent autosomal monosomies in MK involve the chromosomes 7, 5, 17 and 18\\\hline
 \textit{add5qdel5q} & A portion of chromosome inserted into or deleted from long arm (q) of chromosome 5\\\hline
 \textit{add9pdel9p} & A portion of chromosome inserted into or deleted from short arm (p) of chromosome 9\\\hline
 \textit{add12pdel12p}  & A portion of chromosome inserted into or deleted from short arm (p) of chromosome 12\\\hline
  \textit{add14q32} & A portion of chromosome inserted into 14q32\\\hline
 \textit{del7qm7} & A portion of chromosome deleted from long arm (q) of chromosome 7 or \\
                                & missing a copy of chromosome 7\\\hline
 \textit{m17i17qdel17p} & Missing a copy of chromosome 17 or  isochromosome 17 or \\
                                             & a portion of chromosome deleted from short arm (p) of chromosome 17\\\hline
 \textit{add7pi7q} & A portion of chromosome deleted from short arm (p) of chromosome 7 or isochromosome 7\\\hline
 \textit{tratriphyper} & Tetraploid: 4 copies of each chromosome instead of 2 copies or\\
                                      & Near triploidy: 68-80 total chromosomes or\\
                                      & High hiperdiploidy: 51-65 total chromosomes\\\hline
 \textit{lohyperpo} & Low hyperdiploidy: 47-50 total chromosomes or\\
                                  & Low hypodiploidy: 31-39 total chromosomes\\\hline
 \textit{hihyper} & High hyperdiploidy or high hypodiploidy\\\hline
 \textit{complex} & 3 or more distinct abnormalities\\\hline
 \textit{tetra} & Tetraploid: 4 copies of each chromosome instead of 2 copies\\\hline
\end{tabular}}
\end{table}

\end{document}